\documentclass[final,preprint,aps,tightenlines,showpacs]{revtex4}
\usepackage{graphicx}
\usepackage{bm}
\usepackage{dcolumn}


\begin{document}

\bibliographystyle{}


\preprint{\vbox{ \hbox{RUB-TPII-09/99} \hbox{PNU-NTG-01/99}}}
\title{$\Delta S=1,2$ Effective Weak Chiral Lagrangian \\ from the Instanton Vacuum }

\author{
Mario Franz$^{(1)}$
\footnote{email:mariof@tp2.ruhr-uni-bochum.de}, Hyun-Chul
Kim$^{(2)}$ \footnote{email:hchkim@pnu.edu}, and Klaus
Goeke$^{(1)}$ \footnote{email:klaus.goeke@ruhr-uni-bochum.de}
}

\affiliation{
(1) Institute for Theoretical  Physics  II,   P.O. Box 102148,  \\
Ruhr-University Bochum,
 D-44780 Bochum,   Germany \\
(2) Department of Physics, Pusan National University,\\
609-735 Pusan, Republic of Korea
       }
\date{July 2001}

\begin{abstract}
We study the effective weak chiral Lagrangian from the instanton
vacuum.  We incorporate the effective weak Hamiltonian into the
effective low-energy partition function defining the chiral symmetric quark-Goldstone
boson interactions with the momentum-dependent dynamical quark
mass. Employing the derivative expansion and the $1/N_c$ expansion, we derive the
corresponding bosonic weak effective Lagrangian in leading order
with the low energy constants to be used {\em e.g.} in chiral
perturbation theory.  We find that the momentum-dependent dynamical
quark mass plays an essential role in improving the low energy
constants and their ratio $g_{\underline{8}}/g_{\underline{27}}$. \vspace{1cm}

\end{abstract}
\maketitle

\section{Introduction}
Understanding of the nonleptonic decays of light hadrons
in the Standard Model (SM) has been one of the most difficult
issues.  While the SM explains the weak processes involving the change
of strangeness by considering $W$ exchange, the description of
nonleptonic decays of light hadrons is rather complicated
on account of the strong interactions in the low-energy
regime (below $1$ GeV).  The problem is typified by the
$\Delta T = 1/2$ selection rule, best known as the fact that the isosinglet
amplitude of the $K\rightarrow \pi\pi$ decay dominates over
the $T=2$ amplitude by about 22 times.  Despite a great deal of effort
this dominance of the $\Delta T=1/2$ channel over the $\Delta T=3/2$ it
has not been explained in a satisfactory way.  The effective
weak Hamiltonian derived from evolving the simple
$W$-exchange-vertex from a scale of 80 GeV down to
1 GeV~\cite{GaillardLee,AM,Witten,VZS,SVZ,GW,GP,BW,Burasetal}
presents a part of the answer, showing that the perturbative gluons
inherent in the Wilson coefficients enhance the $\Delta T=1/2$ channel.
However, the perturbative gluons alone are not enough to explain
the $\Delta T=1/2$ rule completely.  Thus,
we need to consider other sources of the $\Delta T=1/2$ enhancement.
Since the structure of the light hadrons intimates already the importance of
nonperturbative QCD, it is natural to consider its significance in
describing low energy processes such as $K\rightarrow \pi\pi$ decay.

Recently, we have investigated the effective $\Delta S=1,2$
weak chiral Lagrangian to order ${\cal O}(p^4)$ within the framework of
the chiral quark model ($\chi$QM)~\cite{FranzKimGoeke},
focusing on determining the low energy constants (LECs)
in the effective weak chiral Lagrangian.  However, the
ratio of the $g_8/g_{27}$ obtained from the $\chi$QM deviated from the
phenomenological values obtained in chiral perturbation theory ($\chi$PT).
There was a suggestion to include the gluon condensate which is of order
${\cal O} (\alpha_s N_c)$~\cite{Antonellietal}.
However, Ref.~\cite{Antonellietal} did not consider the ${\cal O}(\alpha_s N_c)^2$ 
corrections which might be of significance at the same extent to the
${\cal O}(\alpha_s N_c)$ corrections.

In the present work, we want to improve our former study~\cite{FranzKimGoeke},
based on the more general effective low-energy QCD partition function
derived from the instanton vacuum which pertains to nonperturbative QCD.
The main feature lies in the fact that the coupling strength of
the chiral symmetric quark-Goldstone interactions is momentum-dependent.
In fact, switching off the momentum dependence of the coupling
and adding an appropriate regularization scheme
will lead to the usual $\chi$QM used in the former
investigation~\cite{FranzKimGoeke}.  Due to the complexity of the formalism
in the present approach we first concentrate on order ${\cal O}(p^2)$
and a part of ${\cal O}(1/N_c)$.  

As far as the large $N_c$ expansion is concerned,
while it explains the strong-interaction sector quantitatively well,
the nonleptonic weak interactions defy any explanation from the strict
large $N_c$ limit.  Because of the fact that the $\Delta T=1/2$ enhancement is
badly underestimated in lowest oder in the large $N_c$ expansion,  
it is inevitable to go beyond the leading order (LO) in
$1/N_c$~\cite{Fukugita,Tadicetal,Chivukula,Bardeenetal,Goity,PichRafael2}.
Moreover, if one expects a large contribution 
from the next-to-leading order (NLO) in $1/N_c$, a convergence problem for 
the $1/N_c$ expansion may occur~\cite{Goity}.  In this case, one has
to consider higher order corrections in $1/N_c$.  Furthermore, since 
there are many different sources of nonleading-order corrections in
$1/N_c$, one has to carefully analyze ${\cal O}(1/N_c)$ corrections.
However, dealing with higher order corrections and all contributions
of ${\cal O}(1/N_c )$ is beyond our work.  
Thus, we will restrict ourselves in this work to a part of 
the $1/N_c$ corrections: The $1/N_c$ diagrams from the quark operators. 

The instanton vacuum elucidates one of the most important low-energy
properties of QCD, {\em i.e.} the mechanism of spontaneous breaking of chiral
symmetry~\cite{DP1,DP2,DP3}.  The Banks-Casher relation tells
that the spectral density $\nu(\lambda)$ of the Dirac operator
at zero modes is proportional to the chiral condensate known as an
order parameter of spontaneous breaking of chiral symmetry:
\begin{equation}
\langle \bar{\psi} \psi\rangle \;=\; -\frac{\pi \nu(0)}{V^{(4)}}.
\end{equation}
The picture of the instanton vacuum provides a good realization of
spontaneous breaking of chiral symmetry. A finite density
of instantons and antiinstantons produces the nonvanishing value of
$\nu(0)$, which triggers the mechanism of chiral symmetry breaking.
The Euclidean quark propagator in the
instanton vacuum acquires the following form with a momentum-dependent
quark mass generated dynamically, identified with the coupling strength
between quarks and Goldstone bosons:
\begin{equation}
S_F (k) \;=\; \frac{\rlap{/}{k}+iM(k)}{k^2 + M^2(k)}
\label{Eq:prop}
\end{equation}
with
\begin{equation}
M(k) \;=\; {\rm const}\cdot \sqrt{\frac{N\pi^2 \bar{\rho}^2}{VN_c}}
F^2 (k\bar{\rho}).
\end{equation}
The ratio $N/V$ denotes the instanton density at equilibrium and
the $\bar{\rho}$ is the average size of the instanton.
The form factor $F(k\bar{\rho})$ is related to the Fourier transform of
the would-be zero fermion mode of individual instantons.
As will be shown later, the momentum dependence of the constituent
quark plays an essential role in improving our previous
results~\cite{FranzKimGoeke} concerning the effective weak chiral Lagrangian.
It is in line with the recent works on the pion wave
functions~\cite{Petrovetal,Praszalowicz:2001wy} and skewed parton distribution~\cite{Skew},
where the momentum-dependent quark mass plays a crucial role as well.

The instanton vacuum induces effective $2N_f$-fermion
interactions~\cite{DP3,DP4}.  For example, it has a type of the
Nambu-Jona-Lasinio model for $N_f=2$ while for $N_f=3$ it exhibits the
't~Hooft determinant~\cite{tHooft}.  Goldstone bosons appear as collective
excitations by quark loops generating a dynamic quark mass.  Eventually
it is found that at low energies QCD is reduced to an interacting
quark-Goldstone boson theory given by the following Euclidean
partition function~\cite{DPP}
\begin{equation}
{\cal Z} = \int {\cal D} \psi {\cal D} \psi^\dagger {\cal D} \pi^a
\exp \left[\int d^4 x \psi^{\dagger \alpha}_{f}
\left(i\rlap{/}{\partial} +
i\sqrt{M(-i\partial)} U^{\gamma_5}
\sqrt{M(-i\partial)}\right)_{fg}
\psi^{\alpha}_{g} \right],
\label{Eq:part}
\end{equation}
where $U^{\gamma_5}$ stands for the pseudo-Goldstone boson:
\begin{equation}
U^{\gamma_5} (x) = U(x) \frac{1+\gamma_5}{2} + U^\dagger (x)
\frac{1-\gamma_5}{2} \;=\;
\exp\left(i\pi^a(x) \lambda^a\gamma_5\right).
\end{equation}
The $\alpha$ is the color index, $\alpha = 1,\cdots, N_c$ and
$f$ and $g$ are flavor indices.  $M(-i\partial) $
is the constituent quark mass
being now momentum-dependent.  It will play a main role in the present
work.  This effective theory of quarks and light Goldstone mesons applies
to quark momenta up to the inverse size of the instanton,
$\bar{\rho}^{-1} \simeq 600$ MeV, which may act as a
scale of the model ($\mu_{\chi{\rm QM}}$).  A merit to derive the
$\chi$QM from the instanton vacuum lies in the fact that the scale of the
model is naturally determined by $\bar{\rho}^{-1}$.  Furthermore,
mesons and baryons can be treated on the same footing
in the $\chi$QM.  For example, the model has been very successful
in describing the properties of the baryons~\cite{review}.

We want to mention that there is a problem related to the gauge
invariance in Eq.(\ref{Eq:part}).  Due to the nonlocality of the
interaction expressed in Eq.(\ref{Eq:part}), the vector and
axial-vector Ward-Takahashi identities are not satisfied, that is,
the conservation of the vector current (CVC) and the partial
conservation of the axial-vector current(PCAC) do not hold.  Thus,
one has to make the effective action gauge-invariant.  A well-known
prescription to solve this problem is to insert the path-ordered
exponent into the effective 
action~\cite{Bos,Holdom,BowlerBirse,PlantBirse,Broniowski,LeeKim}.  
At the scale of the $W$ boson, the charged-current weak interaction
of the quarks is mediated by the $W$ boson only and thus is a product
of two local currents.  However, when the charged-current weak
interaction is scaled down to the hadronic scale, {\em i.e.} $\mu=1$
GeV, the renormalization of the QCD quantum corrections 
make the current-current operator mixed with other different types 
of twist-four local composite operators~\cite{Burasetal}.  
However, in the strict large $N_c$ limit, the contribution of
all other composite operators is suppressed by $1/N_c$ except 
for the current-current one which is the same as that at the $W$
scale.  Thus, it need not be renormalized and the conservation of the
currents must be taken into account.

The outline of the present paper is as follows: In Section 2 we
sketch the instanton-induced chiral quark model, emphasizing in
particular the momentum dependence of the constituent quark mass
and explain how to perform the derivative expansion in the
presence of the momentum-dependent constituent quark mass.
In section 3, we show the basic formalism to obtain the effective
weak chiral Lagrangian starting from the effective weak quark Hamiltonian.
In section 4, we dicuss the large $N_c$ limit in the nonleptonic
weak Hamiltonian and the conserved currents with the nonlocal interaction.  
Section 5 is devoted to the derivation of the 
effective weak chiral Lagrangian to order ${\cal O} (p^2)$ and 
${\cal O}(1/N_c)$.  In Section 6 we
discuss results.  The conclusions and outlook are given in Section 7.
\section{Effective chiral theory from the instanton vacuum}
The effective low-energy QCD partition function in Euclidean space can be
written as
\begin{eqnarray}
{\cal Z}&=& \int {\cal D} \psi {\cal D} \psi^\dagger
{\cal D} \pi^a
\exp \int d^4 x \Big[ \psi^{\dagger \alpha}_{f} (x)
i \rlap{/}{\partial}\psi^{\alpha}_{f}(x)  \cr
&+& i \int \frac{d^4 k d^4 l}{(2\pi)^8} e^{i(k-l)\cdot x}
\sqrt{M(k) M(l)} \psi^{\dagger \alpha}_{f} (k)
\left(U^{\gamma_5}\right)_{fg}
\psi^{\alpha}_{g} (l)\Big],
\label{Eq:Dirac}
\end{eqnarray}
The $M(k)$ is the momentum-dependent constituent quark mass
expressed as follows:
\begin{equation}
\label{Eq:mkfk}
M(k) \;=\; M_0 F^2(k/\Lambda).
\end{equation}
If we choose the $F(k/\Lambda)$ to be constant and add a regularization
({\em e.g.} Pauli-Villars or proper-time),
the partition function becomes just that of the usual $\chi$QM.
The original expression for the $F(k/\Lambda)$~\cite{DP1,DP2},
which is obtained from the Fourier transformation
of the would-be zero fermion mode of individual instantons
with the sharp instanton distribution assumed, is as follows:
\begin{equation}
F(k/\Lambda) \;=\; 2z \left(
I_0 (z) K_1 (z)
- I_1 (z) K_0 (z)
-\frac{1}{z}
I_1 (z) K_1 (z)\right).
\label{Eq:fofk}
\end{equation}
Here $I_0$, $I_1$, $K_0$, and $K_1$ denote
the modified Bessel functions, $z$ is defined as $z=k/(2\Lambda)$ and the cutoff
parameter $\Lambda$ is in this case just the inverse of $\bar{\rho}$.
When $k$ goes to infinity, $F(k/\Lambda)=F(k\bar{\rho})$
becomes
\begin{equation}
F(k\bar{\rho}) \longrightarrow \frac{6}{(k\bar{\rho})^3}.
\end{equation}

Actually, the momentum-dependent quark mass is related to the nonlocal 
regularization in Euclidean space.  There are other ways of
understanding the nonlocal effective interaction without relying on the
instanton vacuum~\cite{Ripka}.  In those methods, the momentum-dependent
quark mass as a regularization appears as delocalizing the quark fields.
So, various types of the $M(k)$ as a regulator with
the regularization parameter $\Lambda$  has been used by different
authors~\cite{Petrovetal,Praszalowicz:2001wy,GolBroRip}.  

Therefore, we will not confine ourselves to the expression given
in Eq.(\ref{Eq:fofk}) but rather try three different types of the
$M(k)$:
\begin{equation}
M(k) = \left \{
\begin{array}{l} \mbox{Eqs.(\ref{Eq:mkfk}, \ref{Eq:fofk})} \\
M_0 \frac{\Lambda^4}{(\Lambda^2 + k^2)^2} \\
 M_0 \exp{\left(-\frac{k^2}{\Lambda^2}\right)}
\end{array} \right. .
\label{Eq:types}
\end{equation}

The $M(k)$ is normalized to $M_0$ at $k=0$.  Originally,
$M(0)$ is taken to be around $350$ MeV corresponding to the values
of the following parameters: $\bar{R}\simeq 1$ fm and
$\bar{\rho}\simeq 0.35$ fm.
The $\bar{R}$ is the average distance between neighboring instantons.
However, we will regard $M_0$ as a free parameter ranging from $200$ MeV
to $450$ MeV and fit for each $M_0$ the parameter $\Lambda$ to
the pion decay constant $f_\pi=93$ MeV.  Figure~1 shows the momentum
dependence of the three different types of $M(k)$ with $M_0=350$~MeV.

Having integrated out the quark fields in the partition function,
we obtain
\begin{equation}
{\cal Z} \;=\; \int {\cal D} \pi^a \exp{\left(-S_{\rm eff}[\pi^a]\right)},
\end{equation}
where the $S_{\rm eff}[\pi^a]$ stands for the effective chiral action:
\begin{equation}
S_{\rm eff} [\pi^a] = -N_c \ln {\rm det} D (U^{\gamma_5}).
\label{Eq:det}
\end{equation}
Here, the $D(U^{\gamma_5})$ is the Dirac operator defined by
\begin{equation}
D \;=\; i\rlap{/}{\partial} + i\sqrt{M(-i\partial)} U^{\gamma_5}
\sqrt{M(-i\partial)}.
\end{equation}
The Dirac operator is not Hermitian, so that it is useful to
divide the effective action into the real and imaginary parts:
\begin{eqnarray}
{\rm Re} S_{\rm eff} &=& \frac12 \left(S_{\rm eff} + S_{\rm eff}^*\right)
\;=\; -\frac12 N_c \ln {\rm det} \left[D^\dagger D\right],\cr
i{\rm Im} S_{\rm eff} &=& \frac12 \left(S_{\rm eff} - S_{\rm eff}^*\right)
\;=\; -\frac12 N_c \ln {\rm det} \left[D /D^\dagger\right].
\end{eqnarray}
In order to calculate the real part, we substract
the vacuum part and use the derivative expansion. We therefore write
\begin{eqnarray}
{\rm Re} S_{\rm eff}[\pi^a] - {\rm Re} S_{\rm eff}[0]
&=& -\frac{N_c}{2} {\rm Tr} \ln
\left(\frac{D^\dagger D}{D^{\dagger}_{0} D_0} \right) \cr
&=&  -\frac{N_c}{2} \int d^4 x \int \frac{d^4 k}{(2\pi)^4}
e^{-i k x} {\rm tr} \ln
\left(\frac{D^\dagger D}{D^{\dagger}_{0} D_0} \right) e^{i k x} \cr
&=& -\frac{N_c}{2} \int d^4 x \int \frac{d^4 k}{(2\pi)^4}
{\rm tr} \ln
\left(\frac{D^\dagger(\partial \rightarrow \partial + i k)
 D(\partial \rightarrow \partial + i k)}
{D^{\dagger}_{0}(\partial \rightarrow \partial + i k)
D_0(\partial \rightarrow \partial + i k)} \right) \cdot 1.
\label{Eq:Seff}
\end{eqnarray}
Here we used a complete set of plane waves for the calculation of the
functional trace, summed over all states and took the trace in $x$.
'tr' then denotes the usual matrix trace over flavor and Dirac indices.
The rhs. of Eq.(\ref{Eq:Seff}) can now be expanded
in powers of the derivatives of the pseudo-Goldstone boson  fields,
$\rlap{/}{\partial} U^{\gamma_5}$, and of $2ik \cdot \partial+\partial^2$.
After some manipulation using
\begin{eqnarray}
\lefteqn{
D^\dagger(\partial \rightarrow \partial + i k)
 D(\partial \rightarrow \partial + i k) =
 \left(i\rlap{/}{\partial} -\rlap{/}{k}
-i\sqrt{M(-i\partial + k)}U^{-\gamma_5} \sqrt{M(-i\partial + k)}
\right) } \cr && \times
\left(i\rlap{/}{\partial} -\rlap{/}{k}
+i\sqrt{M(-i\partial + k)}U^{\gamma_5} \sqrt{M(-i\partial + k)}
\right)  \cr
&=& -\partial^2 + k^2 -2ik \cdot \partial +
\sqrt{M(-i\partial + k)}U^{-\gamma_5}
M(-i\partial + k) U^{\gamma_5} \sqrt{M(-i\partial + k)}
\cr &-& \sqrt{M(-i\partial + k)} \left(\rlap{/}{\partial}U^{\gamma_5}
\right)\sqrt{M(-i\partial + k)}\cr
&=& -\partial^2 + k^2 -2i k\cdot \partial + M^2 - M
\left(\rlap{/}{\partial} U^{\gamma_5} \right) - 2i M\tilde{M'}k_\alpha
\left[ 2\partial_\alpha + U^{-\gamma_5} \left(\partial_\alpha
U^{\gamma_5} \right) \right] \cr
&-& M\tilde{M'} \left[2\partial^2 + U^{-\gamma_5} \left(\partial^2 U^{\gamma_5}
\right) + 2 U^{-\gamma_5} \left(\partial_\alpha U^{\gamma_5}\right)
\partial_\alpha \right] \cr
&-& 2 M\tilde{M''} k_\alpha k_\beta \left[2\partial_\alpha \partial_\beta
+ U^{-\gamma_5} \left(\partial_\alpha\partial_\beta U^{\gamma_5}
\right) + 2 U^{-\gamma_5} \left(\partial_\alpha U^{\gamma_5}
\right) \partial_\beta\right] \cr
&-& 2\tilde{M'}^{2} k_\alpha k_\beta \left[\left(\partial_\alpha U^{-\gamma_5}
\right)\left(\partial_\beta U^{\gamma_5} \right) +
U^{-\gamma_5} \left(\partial_\alpha \partial_\beta U^{\gamma_5}
\right)
+ 2 U^{-\gamma_5}
\left(\partial_\alpha U^{\gamma_5}\right) \partial_\beta +
2\partial_\alpha \partial_\beta \right] \cr
&+& i \tilde{M'} k_\alpha \Big[\left(\partial_\alpha \rlap{/}{\partial}
U^{\gamma_5}\right) + 2\left(\rlap{/}{\partial} U^{\gamma_5}\right)
\partial_\alpha \Big] + {\cal O}(\partial^3) ,
\label{Eq:DD} \\
&D^{\dagger}_{0}&(\partial \rightarrow \partial + i k)
D_0(\partial \rightarrow \partial + i k) =
 -\partial^2 + k^2 + M^2 - 2ik \cdot \partial \cr &&
- 4i M\tilde{M'} k\cdot \partial - 4 \tilde{M'}^{2}
k_\alpha k_\beta\partial_\alpha
\partial_\beta
- 2 M\tilde{M'} \partial^2  - 4 M\tilde{M''} k_\alpha k_\beta
\partial_\alpha \partial_\beta + {\cal O}(\partial^3),
\end{eqnarray}
where
\begin{equation}
\label{Eq:Ms}
M=M(k), \;\; \tilde{M'} = \frac{1}{2k} \frac{dM(k)}{dk},
\;\;\tilde{M''}
=\frac{1}{4k^3}\left(\frac{d^2 M(k)}{dk^2} k
- \frac{d M(k)}{dk}\right) ,
\end{equation}
we obtain eventually the effective chiral action
with the momentum-dependent quark mass:
\begin{equation}
\label{Eq:SeffFpi}
{\rm Re} S_{\rm eff}[\pi^a] - {\rm Re} S_{\rm eff}[0]
\;=\; \frac{f^2_{\pi}}{4} \int d^4 x \left\langle L_\mu L_\mu\right\rangle.
\end{equation}
In Eq.(\ref{Eq:SeffFpi}) $\langle \rangle$ denotes the flavor trace,
$L_\mu = i U^\dagger \partial_\mu U$ is a Hermitian $N_f \times N_f$ matrix
and $f_{\pi}$ denotes the pion decay constant:
\begin{equation}
f_{\pi}^2 \;=\; 4 N_c \int \frac{d^4 k}{(2\pi)^4}
\frac{M^2(k) - \frac12 M(k) M'(k)k
+ \frac14 M'^2(k) k^2}{(k^2 +M^2(k))^2}
\label{Eq:fpiq}
\end{equation}
with
\begin{equation}
M'(k) \;=\; \frac{dM(k)}{dk}.
\end{equation}
When we switch off the momentum dependence of the constituent quark mass,
we end up with the former expression for $f^2_{\pi}$~\cite{FranzKimGoeke}:
\begin{equation}
f_{\pi}^2 \;=\; 4 N_c^{2}\int \frac{d^4 k}{(2\pi)^4}
\frac{M^2}{(k^2 +M^2)^2}, \; M=\mbox{ const}.
\end{equation}
which is logarithmically divergent.  The quark condensate, which is
just the trace of the quark propagator given by Eq.(\ref{Eq:prop}),
is written by
\begin{equation}
\langle \bar{\psi}\psi\rangle_M\;=\;
-i\langle \psi^\dagger\psi\rangle_E\;=\;
-4 N_c \int \frac{d^4 k}{(2\pi)^4}
\frac{M(k)}{k^2 +M(k)^2}.
\label{Eq:quarkcon}
\end{equation}
The subscripts $M$ and $E$ stand for Minkowski and Euclidean space,
respectively.  The gluon condensate is expressed as follows~\cite{DP2}:
\begin{equation}
\left\langle\frac{\alpha_s}{\pi}G^a_{\mu\nu}G^{a\mu\nu}\right\rangle
\;=\; 32 N_c \int \frac{d^4 k}{(2\pi)^4}
\frac{M^2(k)}{k^2 +M^2(k)}.
\label{Eq:gluoncon}
\end{equation}

It is already known that the imaginary part of the effective chiral action
is identical to the Wess-Zumino-Witten action~\cite{WZ,Witten3}
with the correct coefficient, which arises from the derivative expansion
of the imaginary part to order ${\cal O}(p^5)$~\cite{DPP,AF,Chan,DSW}.
An appreciable merit of using the momentum-dependent
quark mass as a regulator was already pointed out by
Ball and Ripka~\cite{BallRipka}.  The momentum-dependent quark mass
provides a consistent regularization of the effective action given in
Eq.(\ref{Eq:det}) in which its real and imaginary parts
are treated on the same footing and thus pertinent observables
such as anomalous decays $\pi^0\rightarrow 2\gamma$ are safely recovered
even if $M(k)$ acts as a regulator.
\section{Effective weak chiral action}
In this section, we will show how the effective weak Hamiltonian
is incorporated into the present framework. The effective weak
chiral action $S^{\Delta S = 1,2}_{\rm eff}[\pi^a]$ can be written as follows:
\begin{equation}
\exp{\left(- S^{\Delta S = 1,2}_{\rm eff}\right)} \;=\; \int {\cal
D} \psi {\cal D} \psi^\dagger \exp \left[\int d^4 x
\left(\psi^{\dagger} D \psi - {\cal H}^{\Delta S = 1,2}_{\rm
eff}\right) \right]. \label{Eq:partw}
\end{equation}
Here the effective weak quark Hamiltonian
${\cal H}^{\Delta S = 1}_{\rm eff}$ consists of ten four-quark operators :
\begin{equation}
{\cal H}^{\Delta S = 1}_{\rm eff}
 \;=\; -\frac{G_F}{\sqrt{2}} V_{ud} V^*_{us}
\sum_i c_i (\mu) {\cal Q}_i (\mu) + {\rm h.c.} .
\label{eq:ham}
\end{equation}
The $G_F$ is the well-known Fermi constant and $V_{ij}$ denote
the Cabibbo-Kobayashi-Maskawa(CKM) matrix elements.
The $\tau$ is their ratio
given by $\tau = - V_{td} V^*_{ts}/V_{ud} V^*_{us}$.
The Wilson coefficients $c_i(\mu)$ are defined as
$c_i (\mu) = z_i (\mu) + \tau y_i (\mu)$.
The functions $z_i (\mu)$ and $y_i (\mu)$ are the scale-dependent
Wilson coefficients given at the scale of the $\mu$. The
$z_i (\mu)$ represent the $CP$-conserving part,
while $y_i (\mu)$ stand for the $CP$-violating one.
The four-quark operators ${\cal Q}_i$ contain the dynamic information
of the weak transitions, being constructed by
integrating out the vector bosons $W^{\pm}$ and $Z$ and
heavy quarks $t$, $b$, and $c$.
The four-quark operators~\cite{Burasetal} are given by
\begin{eqnarray}
{\cal Q}_1  & = & -4 \left( {s}^\dagger_{\alpha}
\gamma_{\mu} P_L u_{\beta} \right)
     \left( {u}^\dagger_{\beta} \gamma_{\mu} P_L  d_{\alpha} \right)
, \;\;\;
{\cal Q}_2   =  -4 \left( {s}^\dagger_{\alpha}
\gamma_{\mu} P_L u_{\alpha} \right)
    \left( {u}^\dagger_{\beta} \gamma_{\mu} P_L d_{\beta}  \right)
,\label{Eq:qq}
 \\
{\cal Q}_3 & = & -4 \left( {s}^\dagger_{\alpha} \gamma_{\mu}
P_L d_{\alpha} \right)
  \sum_{q=u,d,s} \left( {q}^\dagger_{\beta} \gamma_{\mu} P_L  q_{\beta} \right)
, \;\;\;
{\cal Q}_4  =  -4 \left( {s}^\dagger_{\alpha}
\gamma_{\mu} P_L d_{\beta} \right)
  \sum_{q=u,d,s} \left( {q}^\dagger_{\beta} \gamma_{\mu} P_L q_{\alpha} \right)
, \cr
{\cal Q}_5 & = & -4 \left( {s}^\dagger_{\alpha}
\gamma_{\mu} P_L d_{\alpha} \right)
  \sum_{q=u,d,s} \left( {q}^\dagger_{\beta} \gamma_{\mu} P_R q_{\beta} \right)
,  \;\;\;
{\cal Q}_6   =  -4 \left( {s}^\dagger_{\alpha}
\gamma_{\mu} P_L d_{\beta} \right)
  \sum_{q=u,d,s} \left( {q}^\dagger_{\beta} \gamma_{\mu} P_R q_{\alpha} \right)
, \cr
{\cal Q}_7 & = & -6 \left( {s}^\dagger_{\alpha}
\gamma_{\mu} P_L d_{\alpha} \right)
 \sum_{q=u,d,s}
\left( {q}^\dagger_{\beta} \hat{Q} \gamma_{\mu} P_R q_{\beta} \right)
, \;\;\;
{\cal Q}_8  =  -6 \left( {s}^\dagger_{\alpha}
\gamma_{\mu} P_L d_{\beta} \right)
 \sum_{q=u,d,s}
\left( {q}^\dagger_{\beta} \hat{Q} \gamma_{\mu} P_R q_{\alpha} \right)
, \cr
{\cal Q}_9 & = & -6 \left( {s}^\dagger_{\alpha}
\gamma_{\mu} P_L d_{\alpha} \right)
  \sum_{q=u,d,s}
\left( {q}^\dagger_{\beta} \hat{Q} \gamma_{\mu} P_L q_{\beta}  \right)
, \;\;\;
{\cal Q}_{10}  =  -6 \left( {s}^\dagger_{\alpha}
\gamma_{\mu} P_L d_{\beta} \right)
  \sum_{q=u,d,s}
\left( {q}^\dagger_{\beta} \hat{Q} \gamma_{\mu} P_L q_{\alpha}  \right) ,
\nonumber
\end{eqnarray}
where $P_{L,R} = \frac{1}{2} \left( 1 \pm \gamma_5 \right)$ are the chiral
projection operators and $\hat{Q}=\frac13 {\rm diag}(2,-1,-1)$ denote the
quark charge matrix. The ${\cal Q}_1$ and ${\cal Q}_2$ come from the
current-current diagrams, while ${\cal Q}_3$ to ${\cal Q}_6$~\cite{SVZ,GW,GP}
and ${\cal Q}_7$ to ${\cal Q}_{10}$~\cite{BW}
are induced by QCD penguin and electroweak penguin
diagrams, respectively.  Note that only seven operators in Eq.(\ref{Eq:qq})
are independent.  For example, we can express
${\cal Q}_4$, ${\cal Q}_9$, and ${\cal Q}_{10}$ as follows:
\begin{equation}
{\cal Q}_4 = - {\cal Q}_1 + {\cal Q}_2 + {\cal Q}_3,\;\;\;
{\cal Q}_9 = \frac12 \left(3{\cal Q}_1 - {\cal Q}_3\right),\;\;\;
{\cal Q}_{10} = {\cal Q}_2 +
\frac12 \left({\cal Q}_1 - {\cal Q}_3\right).
\end{equation}
Under the chiral transformation ${\rm SU}(3)_L \times {\rm SU(3)}_R$
the four-quark operators $Q_2 - Q_1 ,{\cal Q}_{3,4,5,6}$ transform like
$(8_L, 1_R)$.  The ${\cal Q}_1+ {\cal Q}_2, {\cal Q}_{9,10}$ transform
like the combination of $(8_L, 1_R)$ and $(27_L, 1_R)$,
while the ${\cal Q}_{7,8}$ transform like $(8_L, 8_R)$.

 The $\Delta S = 2$ effective 
weak Hamiltonian is expressed as~\cite{GilmanWise2,BJW,HN1,HN2}
\begin{equation}
{\cal H}^{\Delta S = 2}_{\rm eff} \;=\;-\frac{G^2_{F} M^2_{W}}{16\pi^2}
{\cal F} \left(\lambda_c,\lambda_t,m^2_{c},m^2_{t},M^2_{W}\right)b(\mu) 
{\cal Q}_{\Delta S = 2} (\mu) + \mbox{h.c.}
\end{equation}
with
\begin{equation}
{\cal F} \;=\;\lambda^2_{c} \eta_1  S \left(\frac{m^2_{c}}{M^2_{W}}\right) +
\lambda^2_{t} \eta_2  S \left(\frac{m^2_{t}}{M^2_{W}}\right)
+ 2 \lambda_{c}\lambda_{t} \eta_3  S 
\left(\frac{m^2_{c}}{M^2_{W}},\frac{m^2_{t}}{M^2_{W}}\right)
\end{equation}
and the parameters $\lambda_q=V_{qd}V^{*}_{qs}$ denote the 
pertinent relations of the CKM matrix elements with $q=u,c,t$.  The functions
$S_{i}$ are the Inami-Lim functions~\cite{InamiLim,GilmanWise2,BBH},
being obtained by integrating over electroweak loops
and describing the $|\Delta S| = 2$ transition amplitude in the absence of 
strong interactions.  The $b(\mu)$ is again the corresponding
scale-dependent Wilson coefficient.  The coefficients $\eta_i$ represent the short-distance QCD 
corrections split off from the $b(\mu)$~\cite{HN2}.  The four-quark operator 
${\cal Q}_{\Delta S = 2}$ is written as
\begin{equation}
{\cal Q}_{\Delta S = 2} \;=\; -4 \left(s^{\dagger}_\alpha
\gamma_\mu P_L d_\alpha\right)\left(s^{\dagger}_\beta \gamma_\mu P_L
d_\beta\right). 
\end{equation}

Since the Fermi constant $G_F$ is very small,
one can expand Eq.(\ref{Eq:partw}) in powers of the $G_F$ and keep the
lowest order only.  Then we can obtain the effective weak chiral
Lagrangian
\begin{equation}
{\cal L}^{\Delta S = 1,2}_{\rm eff} \;=\; -\frac{1}{\cal Z} \int
{\cal D} \psi {\cal D} \psi^\dagger {\cal H}^{\Delta S = 1,2}_{\rm
eff} \exp \left[\int d^4 x \psi^\dagger D \psi\right].
\label{Eq:part1}
\end{equation}

If you write a generic operator for the four-quark operator
${\cal Q}_i$ for a given $i$ in Euclidean space such as
\begin{equation}
{\cal Q}_i(x) \;=\; -\psi^\dagger (x) \gamma_\mu {P}_{R,L} \Lambda_1
\psi(x) \psi^\dagger (x) \gamma_\mu {P}_{R,L} \Lambda_2 \psi(x),
\end{equation}
where ${\Lambda}_{1,2}$ denote the flavor
spin operators, then we can calculate the vacuum expectation
value (VEV) of ${\cal Q}_i(x)$ as follows:
\begin{eqnarray}
\langle {\cal Q}_i \rangle &=& \frac{1}{\cal Z}\int {\cal D} \psi {\cal D}
\psi^\dagger {\cal Q}_i(x) \exp \left[\int d^4 z \psi^\dagger D \psi\right]
\nonumber \\
&=& \frac{1}{\cal Z}\int d^4y \int \frac{d^4 k}{(2\pi)^4}
e^{ik(x-y)}
\frac{\delta}{\delta J^{(1)}_{\mu} (x)} \frac{\delta}{\delta J^{(2)}_\mu (y)}
\nonumber \\
&& \times
\exp \left[\int d^4z \left \langle z \left| {\rm tr} \ln \tilde{D}
(J_1(z),J_2(z))\right| z\right\rangle \right]_{J_{1}=J_{2}=0} \nonumber \\
&=& L^{(1)}_i \;+\; L^{(2)}_i.
\end{eqnarray}
Here, $\tilde{D}$ is
\begin{equation}
\tilde{D}(J_1(z),J_2(z)) \;=\; D + J^{(1)}_{\alpha}(z) \gamma_\alpha
{P}_{R,L} \Lambda_1
+ J^{(2)}_{\beta} (z) \gamma_\beta {P}_{R,L} \Lambda_2.
\end{equation}
The $L^{(1)}_i$ and $L^{(2)}_i$ are given by
\begin{eqnarray}
L^{(1)}_i &= & -N^{2}_c  {\rm tr} \left[\left\langle x\left|
\frac{1}{D} \gamma_\mu
{P}_{R,L} \Lambda_1 \right |x \right \rangle \left\langle x \left |
\frac{1}{D}\gamma_\mu
{P}_{R,L} \Lambda_2 \right |x \right \rangle \right]_i
+ {\cal O}\left( N_c \right) \nonumber \\
&=& -N_c^2 {\rm tr} \left[ {(A_1)}_\mu {(A_2)}_\mu
\right]_i
+ {\cal O}\left( N_c \right) \hspace{19mm} i=1,4,6,8,10 \; ,
\label{Eq:wea1}\\
L^{(2)}_i &= & N^{2}_c {\rm tr} \left[ \left \langle x\left |
\frac{1}{D}\gamma_\mu
{P}_{R,L} \Lambda_1 \right | x \right
\rangle \right]{\rm tr} \left[\left \langle x \left |\frac{1}{D}
\gamma_\mu
{P}_{R,L} \Lambda_2 \right | x \right \rangle \right]_i
+ {\cal O}\left( N_c \right) \nonumber \\
&=& N_c^2 {\rm tr} \left[ {(A_1)}_{\mu} \right]
 {\rm tr} \left[  {(A_2)}_{\mu} \right]_i
+ {\cal O}\left( N_c \right) \hspace{10mm}
i=2,3,5,7,9 \;,
\label{Eq:wea2}
\end{eqnarray}
where $\Lambda_{1,2}$ are the corresponding flavor matrices.
Applying the derivative expansion to the operators ${(A_{1,2})}_\mu$ up to 
order ${\cal O}(\partial^2)$ they can be written as
\begin{eqnarray}
{(A_{1,2})}_{\mu} &=&  \sum_{n=0}^{\infty}
\int {d^4k \over (2 \pi)^4} \;
\left(\frac{1}{k^2+M^2(k)}\right)^{n+1}\cr
&\times& \Big[ 2i\left(1+2M\tilde{M'}\right)k_\alpha \partial_\alpha
+ M(\rlap{/}{\partial} U^{\gamma_5})
+ 2i M\tilde{M'} k_\alpha U^{-\gamma_5} (\partial_\alpha U^{\gamma_5})\cr
&+& \left(1+2M\tilde{M'}\right) \partial^2 + M\tilde{M'}\left(
U^{-\gamma_5}(\partial^2 U^{\gamma_5}) + 2U^{-\gamma_5}
(\partial_\alpha U^{\gamma_5})\partial_\alpha \right) \cr
&+& 4 (M\tilde{M''} + \tilde{M'}^2)k_\alpha k_\beta \left(\partial_\alpha \partial_\beta  + U^{-\gamma_5}
(\partial_\alpha  U^{\gamma_5})\partial_\beta +\frac12
U^{-\gamma_5} (\partial_\alpha \partial_\beta U^{\gamma_5})\right)\cr
&+& 2 \tilde{M'}^2 k_\alpha k_\beta (\partial_\alpha  U^{-\gamma_5})
(\partial_\beta U^{\gamma_5})
- i \tilde{M'} k_\alpha
\left((\partial_\alpha \rlap{/}{\partial} U^{\gamma_5})
+ 2(\rlap{/}{\partial} U^{\gamma_5})\partial_\alpha \right)\Big]^n\cr
&\times& \left[-\rlap{/}{k} - i M U^{-\gamma_5}
- \tilde{M'}k_\gamma (\partial_\gamma U^{-\gamma_5})  +\frac{i}{2} \tilde{M'}
(\partial^2 U^{-\gamma_5}) \right.\cr
&+&\left. i \left(\tilde{M''} - \frac12 \frac{\tilde{M'}^2}{M}
\right)k_\gamma k_\delta
 (\partial_\gamma \partial_\delta  U^{-\gamma_5})\right]\gamma_\mu P_{\rm L,R}
\Lambda_{1,2},
\label{Eq:A12}
\end{eqnarray}
where $M$, $\tilde{M'}$ and $\tilde{M''}$ are the $k$-dependent functions
defined in Eq.(\ref{Eq:Ms}).
Before we proceed to determine the low energy constants for the
effective weak chiral Lagrangian, we want to discuss the role of the
large $N_c$ limit and related problems of the current conservation.

\section{The $1/N_c$ expansion and current conservation}
The large $N_c$ expansion enters into two different places:
Wilson coefficients and four-quark operators.  The $N_c$ dependence of
the $6\times 6$ anomalous dimension matrix $\gamma$ in the basis of $Q_{1,2,3,4,5,6}$ 
was shown by Ref.~\cite{GW,Bardeenetal}.  The operators
$Q_{7,8,9,10}$ are not mixed with $Q_{1,2,3,4,5,6}$~\cite{Burasetal},
since the LO and NLO anomalous dimensions $\gamma_{ij}$, $i=1,\cdots
6$, $j=7,\cdots 10$ are zero.  Up to ${\cal O} (1/N_c)$, we write the
matrix of the anomalous dimension~\cite{GW,Bardeenetal,Goity} as 
\begin{equation}
  \gamma_{ij} \;=\; \frac12 \frac{\alpha_{\rm s}}{\pi} \Gamma_{ij}
\end{equation}
with
\begin{equation}
\Gamma_{ij} \;=\; \left(
  \begin{array}{cccccc}
0 & 3 & 0 & 0 & 0 & 0 \\
3 & 0 & 0 & \frac13 & 0 & \frac13 \\
0 & 0 & 0 & \frac{11}{3} & 0 & \frac23 \\
0 & 0 & 3 & \frac{N_f}{3} & 0 & \frac{N_f}{3} \\
0 & 0 & 0 & 0 & 0 & -3 \\
0 & 0 & 0 & \frac{N_f}{3} & 0 & \left(\frac{N_f}{3}-3 N_c \right)
 \end{array}\right),
\end{equation}
where $\alpha_{\rm s} (\sim \frac{1}{N_c})$ denotes the strong running coupling constant. 
 In the strict large-$N_c$ limit, all elements vanish except the
anomalous dimension of the $Q_6$ operator $\gamma_{66}$.
It means that the operator $Q_2$ is not mixed with the other operators
$Q_{i} (i\neq 2)$ and $Q_6$ is only renormalized.  Because of 
$\alpha_{\rm s} \sim \frac{1}{N_c}$, The Wilson coefficients $c_i (i\neq
2)$ go to zero and $c_2$ becomes unity.  Thus, in the strict
large-$N_c$ limit the operator $Q_2$ does not require any
renormalization, so that it restores the original current-current form
of quark charged weak interaction at $\mu=M_W$, {\em i.e.} the
factorized form.

Strictly speaking, the $Q_2$ is the only contribution to the effective
weak chiral Lagrangian in the large $N_c$ limit and other operators
must be treated as the NLO corrections according to the Wilson
coefficients~\cite{Goity}.  However, we will take the weak Hamiltonian
as our starting point more practically without considering the large $N_c$ behavior of
the anomalous dimensions~\cite{Antonellietal}.    

Since the momentum-dependent quark mass presents the nonlocal
interaction between the Goldstone-bosons and quarks, the vector and
axial-vector currents are known to be not conserved.  Making the
effective action gauge-invariant with some approximations, one is able
to derive conserved currents ~\cite{BowlerBirse,PlantBirse,Broniowski,LeeKim}.  
The conserved vector and axial-vector currents in Euclidean space with
the momentum-dependent quark mass are written as
follows~\cite{Broniowski,LeeKim}:
\begin{eqnarray}
  \label{eq:currents}
V_{\mu}^{a} &=& \bar{\psi} \gamma_\mu \lambda^a \psi + i
\langle\bar{\psi} |x\rangle\langle x | \sqrt{M_\mu}
\lambda^a U^{\gamma_5} \sqrt{M} |\psi \rangle 
+ i \langle \bar{\psi} | \sqrt{M}U^{\gamma_5}
\lambda^a  \sqrt{M_\mu} | x\rangle\langle x |\psi \rangle, \\
A_{\mu}^{a} &=& \psi^\dagger \gamma_\mu \gamma_5 \lambda^a \psi + i
\langle\psi^\dagger |x\rangle\langle x | \sqrt{M_\mu}
\gamma_5\lambda^a U^{\gamma_5} \sqrt{M} |\psi \rangle 
- i \langle\psi^\dagger  | \sqrt{M}U^{\gamma_5}
\gamma_5\lambda^a \sqrt{M_\mu} |x\rangle\langle x |\psi
\rangle, \nonumber 
\end{eqnarray}
where $\sqrt{M_\mu}$ denotes $d\sqrt{M} /d p_\mu$.  Last two terms in
$V_{\mu}^a$ and $A_{\mu}^a$ are required so that
the currents can be conserved.  

The pion decay constant $f_\pi$ is related to the following transition
matrix elements:
\begin{equation}
\langle 0 | A_\mu^{a} (x) | \pi^{b} (p) \rangle \;=\; i f_\pi p^\mu
e^{ip\cdot x}\delta^{ab},
  \label{eq:fpim}
\end{equation}
where $A_\mu^{a}$ is defined in Eq.(\ref{eq:currents}).  The
$f_\pi^{2}$ obtained from Eq.(\ref{eq:fpim}) is exactly the same as
Eq.(\ref{Eq:fpiq})~\cite{LeeKim}, which indicates that the PCAC is
well satisfied with the additional nonlocal term in Eq.(\ref{eq:currents}).
With the nonlocal terms turned off, we would end up with the
Pagels--Stokar condition for $f_\pi^{2}$~\cite{Pagels:1979hd}:
\begin{equation}
f_\pi^{2} ({\rm PS}) \;=\; \int { d^4 k \over (2 \pi )^4 }
{ M^2 - \frac{1}{4} M M' k \over {(k^2 + M^2)}^2 }.
\label{eq:PS}
\end{equation}
Although the pion decay constant given in Eq.(\ref{Eq:fpiq}) is the
correct one with the momentum-dependent quark mass,
we want to use the Pagels-Stokar condition for the normalization of
the effective chiral Lagrangian for convenience.  The reason lies in
the fact that by using the Pagels-Stokar condition we need not
consider the additional nonlocal part of the currents in deriving the
VEV of the quark operator, since we obtain the same results as we use the
conserved currents given in Eq.(\ref{eq:currents}) and hence we
reproduce precisely the large $N_c$ results for the $B_K$ factor as well as the LECs.  

\section{Effective weak chiral Lagrangian}
\subsection{Leading order in the $1/N_c$ expansion}
In order of ${\cal O} (p^0)$, only the operator ${\cal Q}_{8}$ has
the nonvanishing term:
\begin{equation}
\langle {\cal Q}_8+{\cal Q}^{\dagger}_8 \rangle_{{\cal O} (p^0)}
\;=\; 48 N^2_{c} {\cal M}^2 \langle U \lambda_6 U^\dagger \hat{Q}\rangle.
\end{equation}
However, since we are mainly interested in the LECs
$g_{\underline{8}}$ and $g_{\underline{27}}$ of the effective weak chiral Lagrangian in LO,
we proceed to calculate the NLO terms, {\em i.e.} those in ${\cal O} (p^2)$ order:
\begin{eqnarray}
\langle {\cal Q}_1+{\cal Q}^{\dagger}_1 \rangle_{{\cal O}
  (p^2)}^{{\cal O} (N_c^2)}
& = & 16 N^2_{c} {\cal K}^2
\left(-\frac25 \langle \lambda_6 L_\mu L_\mu
\rangle + \frac13 t_{jl}^{ik} \langle \lambda_{i}^{j} L_\mu \rangle
\langle \lambda_{k}^{l} L_\mu \rangle\right),  \label{Eq:qa}\\
\langle {\cal Q}_2+{\cal Q}^{\dagger}_2\rangle_{{\cal O} (p^2)}^{{\cal O} (N_c^2)}
& = & 16 N_{c} {\cal K}^2 \left(\frac35 \langle \lambda_6 L_\mu L_\mu
\rangle + \frac13 t_{jl}^{ik} \langle \lambda_{i}^{j} L_\mu \rangle
\langle \lambda_{k}^{l} L_\mu \rangle\right),\\
\langle {\cal Q}_3+{\cal Q}^{\dagger}_3\rangle_{{\cal O} (p^2)}^{{\cal O} (N_c^2)}  & = & 0,\\
\langle {\cal Q}_4+{\cal Q}^{\dagger}_4\rangle_{{\cal O} (p^2)}^{{\cal O} (N_c^2)}
& = & 16 N^2_{c} {\cal K}^2
\langle \lambda_6  L_\mu  L_\mu \rangle,  \\
\langle {\cal Q}_5+{\cal Q}^{\dagger}_5\rangle_{{\cal O} (p^2)}^{{\cal O} (N_c^2)} & = & 0,\\
\langle {\cal Q}_6+{\cal Q}^{\dagger}_6\rangle_{{\cal O} (p^2)}^{{\cal O} (N_c^2)} & = &
64 N^2_{c} {\cal M} \left({\cal P} + {\cal R}\right)
\langle \lambda_6  L_\mu  L_\mu \rangle,  \\
{\langle {\cal Q}_7+{\cal Q}^{\dagger}_7 \rangle}_{{\cal O}(p^2)}^{{\cal O} (N_c^2)} &=&
24 N^2_{c} {\cal K}^2
\langle L_\mu \lambda_6 \rangle \langle R_\mu \hat{Q} \rangle,\\
{\langle {\cal Q}_8+{\cal Q}^{\dagger}_8 \rangle}_{{\cal O}(p^2)}^{{\cal O} (N_c^2)} &=
&
48 N^2_{c} {\cal M} \left({\cal P} + {\cal R}\right)
\Big[
\langle U \lambda_6 \left( \partial_\mu U^\dagger \right)
\left( \partial_\mu U \right) U^\dagger  \hat{Q} \rangle
\cr
&+& \langle  \left( \partial_\mu U \right)
\left( \partial_\mu U^\dagger \right) U \lambda_6 U^\dagger
\hat{Q} \rangle \Big], \\
\langle {\cal Q}_9+{\cal Q}^{\dagger}_9\rangle_{{\cal O} (p^2)}^{{\cal O} (N_c^2)}
& = & 16 N^2_{c} {\cal K}^2
\left(-\frac35 \langle \lambda_6 L_\mu L_\mu \rangle
+ \frac12 t_{jl}^{ik} \langle \lambda_{i}^{j} L_\mu \rangle
\langle \lambda_{k}^{l} L_\mu \rangle\right),\\
\langle {\cal Q}_{10}+{\cal Q}^{\dagger}_{10}\rangle_{{\cal O}
  (p^2)}^{{\cal O} (N_c^2)} & = &
16 N^2_{c} {\cal K}^2
\left(\frac25 \langle \lambda_6 L_\mu L_\mu  \rangle
+ \frac12 t_{jl}^{ik} \langle \lambda_{i}^{j} L_\mu \rangle
\langle \lambda_{k}^{l} L_\mu \rangle\right), 
\label{Eq:qz} \\
\langle {\cal Q}_{\Delta S = 2} + {\cal Q}^{\dagger}_ {\Delta S = 2}
\rangle_{{\cal O} (p^2)}^{{\cal O} (N_c^2)} &=& 16 N^2_{c} {\cal K}^2 
\langle \lambda_6 L_\mu \lambda_6 L_\mu \rangle ,
\label{Eq:s2}
\end{eqnarray}
where $t_{jl}^{ik}$ are the eikosiheptaplet projection 
operators~\cite{FranzKimGoeke,Kambor:1990tz}  and 
the functions ${\cal K}$, ${\cal M}$, ${\cal P}$, and ${\cal R}$
are  expressed as follows:
\begin{eqnarray}
{\cal K} &=&
\int { d^4 k \over (2 \pi )^4 }
{ M^2 - \frac{1}{4} M M' k \over {(k^2 + M^2)}^2 }\;=\; f_\pi^{2}
({\rm PS}), \cr
{\cal M} &=& - \int { d^4 k \over (2 \pi )^4 }
{ M \over k^2 + M^2 }\;=\;\frac{\langle\bar{\psi}\psi\rangle_M}{4N_c},\cr
{\cal P} &=& \int { d^4 k \over (2 \pi )^4 }
\left[ { \frac{{M'}^2}{32 M} \over k^2 + M^2} + { M - \frac{1}{4} M {M'}^2
\over {(k^2 + M^2)}^2 } + {
- \frac12 M k^2 - \frac12 M^2 M' k + \frac14 M^3 {M'}^2
\over {(k^2 + M^2)}^3 } \right]
,\cr
{\cal R} &=& \int { d^4 k \over (2 \pi )^4 }
{ \frac12 M^2 M' k - M^3 \over {(k^2 + M^2)}^3 }.
\end{eqnarray}
Note that the function ${\cal K}$ is just the same as the
Pagels-Stokar condition and the function ${\cal M}$ is directly related to the
quark condensate.  
The $\Delta S=1$ effective weak chiral Lagrangian~\cite{Cronin} is
given in Minkowski space as follows: 
\begin{eqnarray}
{\cal L}^{\Delta S = 1,{\cal O} (p^2)}_{\rm eff} & = &
-\frac{G_F}{\sqrt{2}} V_{ud} V^*_{us} f^4_{\pi}\left(
g_{\underline{8}} {\cal L}_{\underline{8}} 
+ g_{\underline{27}} {\cal L}_{\underline{27}} \right)\cr
&=&
-\frac{G_F}{\sqrt{2}} V_{ud} V^*_{us} f^4_{\pi}\left(
g_{\underline{8}}^{(1/2)} {\cal L}^{(1/2)}_{\underline{8}} \;+\; 
g_{\underline{27}}^{(1/2)}{\cal L}^{(1/2)}_{\underline{27}} \;+\;
g_{\underline{27}}^{(3/2)} {\cal L}^{(3/2)}_{\underline{27}}\right),
\label{Eq:Lcpt2}
\end{eqnarray}
where
\begin{eqnarray}
 {\cal L}_{\underline{8}} &=& \left\langle \lambda_3^{2} L_\mu L^\mu
 \right\rangle \;+\;{\rm h.c.},\cr
{\cal L}_{\underline{27}} &=& \frac23  \left\langle \lambda_1^{2}L_\mu \right\rangle
 \left\langle \lambda_1^{3} L^\mu \right\rangle 
+ \left\langle \lambda_2^{3}L_\mu \right\rangle
 \left\langle \lambda_1^{1} L^\mu \right\rangle \;+\; {\rm h.c.}\cr
&=& \frac19 {\cal L}_{\underline{27}}^{(1/2)} \;+\; \frac59
 {\cal L}_{\underline{27}}^{(3/2)},
\label{Eq:lag1} 
\end{eqnarray}
and 
\begin{eqnarray}
{\cal L}^{(1/2)}_{\underline{8}} &=& \left\langle\lambda_{2}^{3}
L_\mu L^\mu \right\rangle\;+\; {\rm h.c.},\nonumber \\
{\cal L}^{(1/2)}_{\underline{27}} &=& \left\langle \lambda_{1}^{2} L_\mu\right\rangle
\left\langle \lambda_{3}^{1} L^\mu \right\rangle
-\left\langle \lambda_{3}^{2} L_\mu\right\rangle
\left\langle \lambda_{1}^{1} L^\mu \right\rangle - 
5 \left\langle \lambda_{3}^{2} L_\mu\right\rangle
\left\langle \lambda_{3}^{3} L^\mu \right\rangle + {\rm h.c.},
\nonumber \\
{\cal L}^{(3/2)}_{\underline{27}} &=&
\left\langle \lambda_{1}^{2} L_\mu\right\rangle
\left\langle \lambda_{3}^{1} L^\mu \right\rangle
+2\left\langle \lambda_{3}^{2} L_\mu\right\rangle
\left\langle \lambda_{1}^{1} L^\mu \right\rangle
+ \left\langle \lambda_{3}^{2} L_\mu\right\rangle
\left\langle \lambda_{3}^{3} L^\mu \right\rangle + {\rm h.c.}.
\label{Eq:L8L27}
\end{eqnarray}
The $g_{\underline{8}}$ and $g_{\underline{27}}$ are dimensionless
LECs of which the numerical values can be extracted from the $CP$-conserving
$K\rightarrow \pi\pi$ decay rate and the $\Delta T = 1/2$ enhancement
is reflected in these constants.  From the analysis in chiral
perturbation theory with chiral loops
considered~\cite{PichRafael,Kamboretal},
the LECs have the following values
\begin{equation}
  \label{eq:emplec}
|g_{\underline{8}} |_{\rm exp} \simeq 3.6,\;\;
|g_{\underline{27}} |_{\rm exp} \simeq 0.29,\;\;
\frac{|g_{\underline{8}}|_{\rm exp}}{|g_{\underline{27}}|_{\rm exp}}\simeq 12.5,   
\end{equation}
 
Comparing Eq.(\ref{Eq:Lcpt2}) with Eqs.(\ref{Eq:qa}-\ref{Eq:qz}), we determine the values
of the LECs $g_{\underline{8}}$ and $g_{\underline{27}}$ to the LO in the $1/N_c$ expansion:
\begin{eqnarray}
g_{\underline{8}} &=& -\frac25 c_1 + \frac35 c_2 + c_4 -\frac35 c_9 + \frac25 c_{10}
+\frac{64 N_{c}^{2} {\cal M} ({\cal P} + {\cal R})}{f_{\pi}^{4}}c_6,
\nonumber \\
g_{\underline{27}} &=& \frac35 c_1 + \frac35 c_2 + \frac{9}{10} c_9
+\frac{9}{10} c_{10}.
  \label{eq:lecsLO}
\end{eqnarray}
In the strict large $N_c$ limit in which the Wilson coefficient $c_2$
survives only and becomes one, we correctly reproduce the following
large $N_c$ results~\cite{PichRafael}:
\begin{eqnarray}
\left. g_{\underline{8}}  \right|_{N_c\rightarrow \infty} &=&
\frac35,\cr 
\left. g_{\underline{27}} \right|_{N_c\rightarrow \infty} &=&
\frac35.
  \label{eq:lecnc}
\end{eqnarray}

The effective $\Delta S = 2$ weak chiral Lagrangian to order ${\cal
  O}(p^2)$ is derived as:
\begin{equation}
{\cal L}^{\Delta S=2,{\cal O}(p^4)}_{\rm eff} 
\;=\; -\frac{G^2_{F} M^2_{W}}{4\pi^2}
{\cal F} \left(\lambda_c, \lambda_t,m^2_{c}, m^2_{t}, M^2_{W}\right) 
\frac43 f_\pi^{4}  \hat{B}_K 
\langle \lambda_6 L_\mu\rangle\langle \lambda_6 L^\mu\rangle,
\label{eq:delta2}
\end{equation}
where $\hat{B}_K$ is known as the scale-independent $B_K$ factor which is defined as
$B_K=B_K (\mu) b(\mu)$.  The scale-dependent $B_K(\mu)$ is related to the matrix
element for the $\bar{K}^0-K^0$ mixing
\begin{equation}
  \langle \bar{K}^0 | {\cal Q}_{\Delta S = 2} | K^0\rangle \;=\;
  \frac{16}{3} B_K(\mu) f_K^{2} m_K^{2},
\end{equation}
which governs the $\bar{K}^0-K^0$ mixing at short distances. Here, 
$f_K$ and $m_K$ are the mass and decay constant of the neutral kaon, respectively.  
Comparing Eq.(\ref{eq:delta2}) with Eq.(\ref{Eq:s2}) in the chiral
limit ($f_\pi=f_K$), we immediately obtain the $B_K$ factor in the large $N_c$ 
limit~\cite{Gaiser:1981gx,Bardeen:1988vg}:
\begin{equation}
  \label{eq:bklo}
  B_K\;=\; \frac34.
\end{equation}

\subsection{${\cal O}(N_c)$ corrections}
Now, we proceed to add the next-to-order corrections in the large $N_c$
expansion.  Though there are many different origins of the NLO
corrections in the $N_c$ expansion, we are not able to consider all
possible NLO corrections.  Here, we will restrict ourselves the 
$1/N_c$ corrections from the quark-quark operators.  The VEVs of the
quark operators in the NLO are as follows:
\begin{eqnarray}
\langle {\cal Q}_1+{\cal Q}^{\dagger}_1 \rangle_{{\cal O}
  (p^2)}^{{\cal O} (N_c)}
& = & 16 N_{c} {\cal K}^2
\left(\frac35 \langle \lambda_6 L_\mu L_\mu
\rangle + \frac13 t_{jl}^{ik} \langle \lambda_{i}^{j} L_\mu \rangle
\langle \lambda_{k}^{l} L_\mu \rangle\right),  \label{Eq:qanlo}\\
\langle {\cal Q}_2+{\cal Q}^{\dagger}_2\rangle_{{\cal O} (p^2)}^{{\cal O} (N_c)}
& = & 16 N_{c} {\cal K}^2
\left(-\frac25 \langle \lambda_6 L_\mu L_\mu
\rangle + \frac13 t_{jl}^{ik} \langle \lambda_{i}^{j} L_\mu \rangle
\langle \lambda_{k}^{l} L_\mu \rangle\right),\\
\langle {\cal Q}_3+{\cal Q}^{\dagger}_3\rangle_{{\cal O} (p^2)}^{{\cal
  O} (N_c)}  & = & 16 N_{c} {\cal K}^2\langle \lambda_6  L_\mu  L_\mu \rangle,  \\
\langle {\cal Q}_4+{\cal Q}^{\dagger}_4\rangle_{{\cal O} (p^2)}^{{\cal O} (N_c)}
& = & 0,\\ 
\langle {\cal Q}_5+{\cal Q}^{\dagger}_5\rangle_{{\cal O} (p^2)}^{{\cal
  O} (N_c)} & = & 
64 N_{c} {\cal M} \left({\cal P} + {\cal R}\right)
\langle \lambda_6  L_\mu  L_\mu \rangle,  \\
\langle {\cal Q}_6+{\cal Q}^{\dagger}_6\rangle_{{\cal O} (p^2)}^{{\cal O} (N_c)} & = & 0,\\
{\langle {\cal Q}_7+{\cal Q}^{\dagger}_7 \rangle}_{{\cal O}(p^2)}^{{\cal O} (N_c)} &=&
48 N_{c} {\cal M} \left({\cal P} + {\cal R}\right)
\Big[
\langle U \lambda_6 \left( \partial_\mu U^\dagger \right)
\left( \partial_\mu U \right) U^\dagger  \hat{Q} \rangle
\cr
&+& \langle  \left( \partial_\mu U \right)
\left( \partial_\mu U^\dagger \right) U \lambda_6 U^\dagger
\hat{Q} \rangle \Big], \\
{\langle {\cal Q}_8+{\cal Q}^{\dagger}_8 \rangle}_{{\cal
  O}(p^2)}^{{\cal O} (N_c)} &=& 24 N_{c} {\cal K}^2
\langle L_\mu \lambda_6 \rangle \langle R_\mu \hat{Q} \rangle, \\
\langle {\cal Q}_9+{\cal Q}^{\dagger}_9\rangle_{{\cal O} (p^2)}^{{\cal O} (N_c)}
& = & 16 N_{c} {\cal K}^2
\left(\frac25 \langle \lambda_6 L_\mu L_\mu  \rangle
+ \frac12 t_{jl}^{ik} \langle \lambda_{i}^{j} L_\mu \rangle
\langle \lambda_{k}^{l} L_\mu \rangle\right), \\
\langle {\cal Q}_{10}+{\cal Q}^{\dagger}_{10}\rangle_{{\cal O} (p^2)}^{{\cal O} (N_c)} & = &
16 N_{c} {\cal K}^2
\left(-\frac35 \langle \lambda_6 L_\mu L_\mu \rangle
+ \frac12 t_{jl}^{ik} \langle \lambda_{i}^{j} L_\mu \rangle
\langle \lambda_{k}^{l} L_\mu \rangle\right),\\
\label{Eq:qznlo}
\langle {\cal Q}_{\Delta S = 2} + {\cal Q}^{\dagger}_ {\Delta S = 2}
\rangle_{{\cal O} (p^2)}^{{\cal O} (N_c)} &=& 16 N_{c} {\cal K}^2 
\langle \lambda_6 L_\mu \lambda_6 L_\mu \rangle .
\label{Eq:s2nlo}
\end{eqnarray}

The LECs $g_{\underline{8}}$ and $g_{\underline{27}}$
are then obtained as follows:
\begin{eqnarray}
g_{\underline{8}} &=& \left(-\frac25 +\frac{3}{5N_c}\right)c_1 
+ \left(\frac35 -\frac{2}{5N_c} \right)c_2 + \frac{1}{N_c}c_3 + c_4 
+\left(-\frac35 + \frac{2}{5N_c}\right) c_9 + \left(\frac25
  -\frac{3}{5N_c}\right) c_{10} \cr
&&+\frac{64 N_{c} {\cal M} ({\cal P} + {\cal R})}{f_{\pi}^{4}}c_5
+\frac{64 N_{c}^{2} {\cal M} ({\cal P} + {\cal R})}{f_{\pi}^{4}}c_6,
\nonumber \\
g_{\underline{27}} &=& \left(1+\frac{1}{N_c}\right)
\left(\frac35 c_1 + \frac35 c_2 + \frac{9}{10} c_9
+\frac{9}{10} c_{10}\right).
\label{eq:lecsNLO}
\end{eqnarray}
If we turn off the momentum dependence of the $M(k)$ and introduce a regularization
with the cut-off parameter $\Lambda$ for the quark-loop integrals,
we end up with the former results in Ref.~\cite{FranzKimGoeke}:
\begin{eqnarray}
{\cal K} &=& \frac{f_{\pi}^{2}}{4N_c}, \cr
{\cal M} &=& \frac{\langle \bar{\psi} \psi\rangle}{4N_c}, \cr
{\cal P} &=& \frac{f_{\pi}^{2}}{8N_cM} + \frac{M}{64\pi^2}, \cr
{\cal R} &=& -\frac{M}{32\pi^2}.
\end{eqnarray}
Thus, the LECs (\ref{eq:lecsNLO}) are reduced to those with the constant constituent quark
mass~\cite{FranzKimGoeke}:
\begin{eqnarray}
g_{\underline{8}} &=& \left(-\frac25 +\frac{3}{5N_c}\right)c_1 
+ \left(\frac35 -\frac{2}{5N_c} \right)c_2 + \frac{1}{N_c}c_3 + c_4 
+\left(-\frac35 + \frac{2}{5N_c}\right) c_9 + \left(\frac25
  -\frac{3}{5N_c}\right) c_{10} \cr
&&+\left(\frac{2\langle\bar{\psi}\psi\rangle}{N_c f_\pi^{2} M} 
- \frac{\langle\bar{\psi}\psi\rangle M}{4 f_\pi^{4}\pi^2}\right)c_5
+\left(\frac{2\langle\bar{\psi}\psi\rangle}{f_\pi^{2} M} 
- \frac{N_c\langle\bar{\psi}\psi\rangle M}{4 f_\pi^{4}\pi^2}
\right) c_6, \nonumber \\
g_{\underline{27}} &=& \left(1+\frac{1}{N_c}\right)
\left(\frac35 c_1 + \frac35 c_2 + \frac{9}{10} c_9
+\frac{9}{10} c_{10}\right).
  \label{Eq:ratio2}
\end{eqnarray}

From Eq.(\ref{Eq:s2nlo}), the $B_K$ factor with the ${\cal O}(N_c)$
correction becomes 
\begin{equation}
  \label{eq:bknlo}
  B_K \;=\;  \frac34 \left(1 + \frac{1}{N_c}\right),
\end{equation}
which is just one with $N_c = 3$~\cite{Gaillard:1974hs} and agrees with 
the result of $1/N_c$ approach~\cite{Buras:1986yx,Bardeen:1988vg}.
\section{Results and Discussion}
We employed the Wilson coefficients $c_i$ obtained by Buchalla
{\em et al.}~\cite{Burasetal}.  There are three different
renormalization schemes in Ref.~\cite{Burasetal}.  For our best fit, we choose the naive
dimensional regularization (NDR) scheme in this work.   

While in the former calculation based on the usual
chiral quark model~\cite{FranzKimGoeke} three parameters are involved,
namely the pion decay constant, the quark condensate, and the constituent
quark mass, the present results given by Eqs.(\ref{eq:lecsLO},\ref{eq:lecsNLO}) include
four functions ${\cal K}$, ${\cal M}$, ${\cal P}$, and ${\cal R}$.
However, those functions depend on the $M(k)$ containing
the $\Lambda$ and $M_0$, so that we have only two free parameters.
Furthermore, the cut-off parameter $\Lambda$ is fixed by reproducing
the pion decay constant in our calculation.  The value of $M_0$ is
more or less constrained by demanding the calculated quark condensate
(\ref{Eq:quarkcon}) and gluon condensate (\ref{Eq:gluoncon}) to lie inside
the empirical limits of $-(350 \ {\rm MeV})^3 \leq \langle \bar{\psi}\psi
\rangle\leq -(200 \ {\rm MeV})^3$~\cite{Danieletal,Fukugitaetal,DoschNarison}
and $(350 \ {\rm MeV})^4\leq \langle \frac{\alpha_s}{\pi}GG\rangle
\leq (400 \ {\rm MeV})^4$~\cite{BNP,Narison,GPV}.  Thus, we first want
to examine the dependence of the quark and gluon condensates
on the momentum-dependent quark mass (see Eqs.(\ref{Eq:quarkcon},\ref{Eq:gluoncon})).

In Figure 2 we show the dependence of the quark condensate on the
$M_0$ with several different types of the $M(k)$ given
in Eq.(\ref{Eq:types}).  The $M(k)$ by Diakonov and
Petrov~\cite{DP1,DP2} and that of the dipole type produce the very similar
results of the $\langle \bar{\psi}\psi\rangle$~\footnote{The monopole type
of the $M(k)$ is not allowed, since it is not enough to tame the
quadratic divergence like the quark condensate.}, the Gaussian type
brings out noticeably smaller value of the quark condensate than the other two.
From these results one can easily see that the $M(k)$ with stronger
tail produces the larger value of the quark condensate.  The shaded band
depicts the empirical range of the quark condensate.

In Figure 3 we draw the dependence of the gluon condensate on the
$M_0$ in the same manner.  As in the case of the quark condensate,
the dependence on the different types of the $M(k)$ is similar.  
Again, the shaded band represents the empirical range of the gluon condensate.
Thus, examining the dependence of the quark and gluon condensates, one constrains
the range of the value of $M_0$.

In Figures 4 and 5 we present dependence of the LECs
$g_{\underline{8}}$ and $g_{\underline{27}}$ on the dynamical quark mass, with
the NDR scheme employed ($\Lambda_{\bar{MS}}^{(4)}=435$ MeV).
As seen in Eqs.(\ref{eq:lecsLO},\ref{eq:lecsNLO}), the
$g_{\underline{27}}$ does not depend on $M_0$.  On the other hand, the
$g_{\underline{8}}$ shows a strong dependence on the $M_0$. It is due
to the fact that the penguin operator contributes only to the $g_{\underline{8}}$ and 
it brings about three functions containing the $M_0$ and the quark
condensate.  The behavior of the $g_{\underline{8}}$ with different types of $M(k)$ 
is similar to the quark and gluon condensates.  In particular,  
the $g_{\underline{8}}$ goes up drastically below $M_0=250$
MeV.  As a result, the ratio $g_{\underline{8}}/g_{\underline{27}}$
exhibits a similar dependence on the $M_0$ as shown in Fig. 6.  

Figures 7 and 8 depict the contribution of the NLO corrections ${\cal O} (N_c)$ to
the LECs $g_{\underline{8}}$ and $g_{\underline{27}}$.  The ${\cal O}
(N_c)$ contribution to the $g_{\underline{8}}$ is very tiny, while it
does to the $g_{\underline{27}}$ almost $30\%$ and thus it
makes the $g_{\underline{27}}$ deviate from the empirical value.  As
already discussed in Ref.\cite{FranzKimGoeke}, the ${\cal O}(N_c)$ corrections
make the ratio $g_{\underline{8}}/g_{\underline{27}}$ worse
as shown in Fig. 6, compared to the empirical value (dotted line).  We only can
reach the empirical value below around $M_0 = 230$ MeV which is not
preferable according to the proper range of the $M_0$
from the quark and gluon condensates.  

Now, we want to analyze the contributions to the LECs from
the different quark operators $Q_i$ given in Eq.(\ref{Eq:qq}).  These
considerations are basically independent of the actual form of $M(k)$.
Since the penguin operator $Q_6$ contributes only to the octet part
of the coupling, it contributes directly to the $\Delta T= 1/2$ enhancement.
In fact, the appearance of this octet penguin operator indicates already that
the effect of perturbative gluons is of utmost importance in explaining the
$\Delta T= 1/2$ rule.  Figure 9 shows each contribution of the quark
operators $Q_i$, starting from the $Q_2$ which has the same form of the
bare operator at the scale of the $M_W$.  The operator $Q_2$ has the largest
Wilson coefficient in any renormalization scheme.  The $Q_1$ takes the second
largest Wilson coefficient.  As drawn in Fig. 9, the LEC $g_{\underline{8}}$
is much underestimated with the operators $Q_1$ and $Q_2$ only and furthermore
they are stable to the change of the $M_0$.
Adding the penguin operator $Q_6$ brings the $g_{\underline{8}}$ up dramatically and
its dependence on the $M_0$ is also very noticeable.  The other contributions
are negligibly tiny.  Hence, the penguin operator $Q_6$ plays an essential role
in enhancing the octet coupling.  Actually, considering the fact that
the $M(k)$ is pertinent to the zero mode of individual instantons,
the strong dependence of the penguin operator on the $M$ indicates
the importance of the effect of a part of nonperturbative gluons
in describing the nonleptonic kaon decays.  

In the case of the $g_{\underline{27}}$,  the operators $Q_1$ and
$Q_2$ are dominant and the Wilson coefficients $c_9$ and $c_{10}$ are
negligibly small.    
\section{Summary and Conclusions}
In the present work we have investigated the effective weak chiral Lagrangian
for $\Delta S=1,2$ concentrating on the calculation of $g_{\underline{8}}$ and
$g_{\underline{27}}$.  We used the chiral quark model from the instanton vacuum as
framework and incorporated the weak interaction by the effective Hamiltonian
of Buchalla, Buras, and Lautenbacher~\cite{Burasetal}.  The calculation
has been done in a first step to order ${\cal O}(p^2)$ and to LO and NLO
in the $1/N_c$ expansion.  In contrast to the previous work by Ref.~\cite{FranzKimGoeke},
we used a momentum-dependent constituent quark mass $M(k)$
as it arises from the instanton vacuum, which makes $g_{\underline{8}}$
be closer to its corresponding empirical value.  However, the
$g_{\underline{27}}$ is untouched by the present work, so that it is
not at all improved, since it is independent of $M$.    

For their ratio $g_{\underline{8}}/g_{\underline{27}}$, this simple
feature enlarged the values by about $50$~$\%$ to $100$~$\%$ and hence
very much improved the theoretical values compared with the
empirical ones.  In fact we tried various shapes of $M(k)$.
However, the best values were obtained for the $M(k)$ from the instanton
model of Diakonov and Petrov~\cite{DP1,DP2} and $M_0\simeq 230$ MeV.  
Altogether the empirical value of $g_{\underline{8}}/g_{\underline{27}}$
with the $M_0$ constrained to the quark and gluon condensates 
is still underestimated and one is still away from
an explanation of the $\Delta T=1/2$ rule.  
However, the conclusion is very clear: If one wishes to derive the effective weak Lagrangian from
the instanton vacuum of QCD by means of the chiral quark model the use of
a momentum-dependent quark mass seems to be indispensable.

\section*{Acknowledgments}
The work is supported in part by COSY, DFG, and BMBF.  The work of HCK
is supported by the Korean Research Foundation (KRF\--2000\--015\--DP0069). 

\newpage

\centerline{\large \bf Figures}

\vspace{0.8cm}
\centerline{\includegraphics[width=15cm,height=12.0cm]{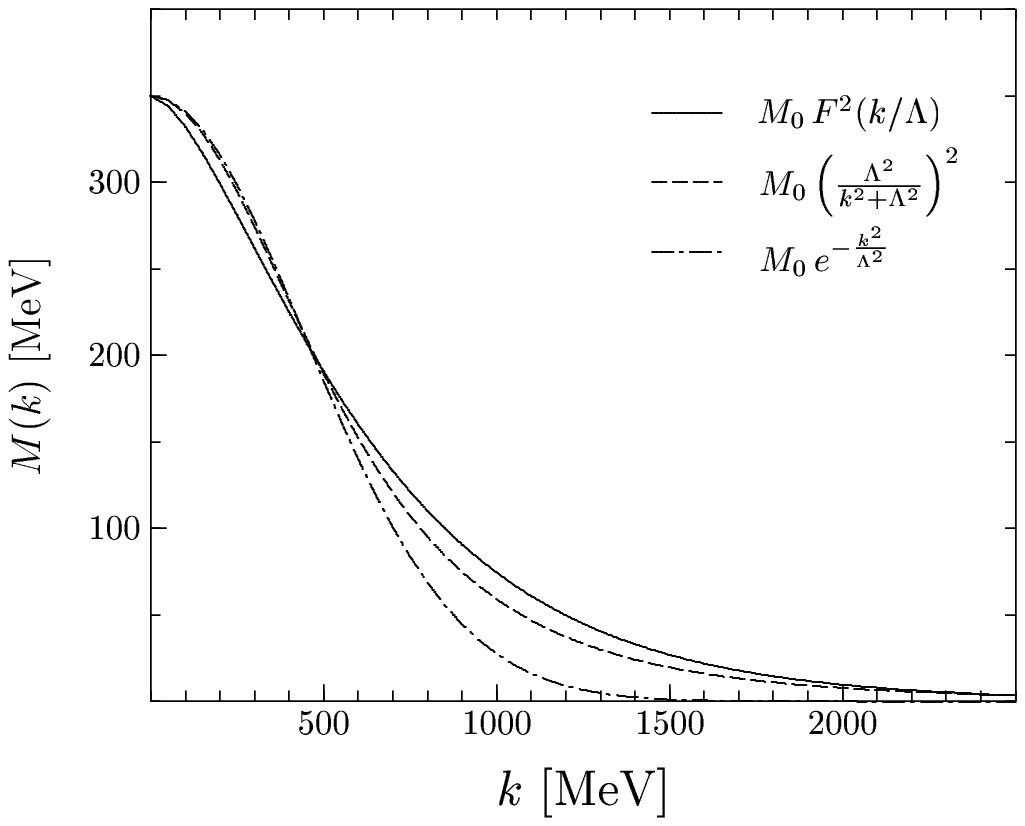}}

\vspace{0.8cm}
\noindent {\bf Fig.1}: Momentum dependence of the constituent quark mass
$M(k)$. The solid curve denotes the original $M(k)=M_0F^2 (k/\Lambda)$ by
Diakonov and Petrov, while the long-dashed one stands for the dipole-type
$M(k)$ and the dot-dashed one depicts the Gaussian one.
The $\Lambda$ has been fitted to reproduce $f_\pi=93$ MeV.

\vspace{0.8cm}
\centerline{\includegraphics[width=15cm,height=12.0cm]{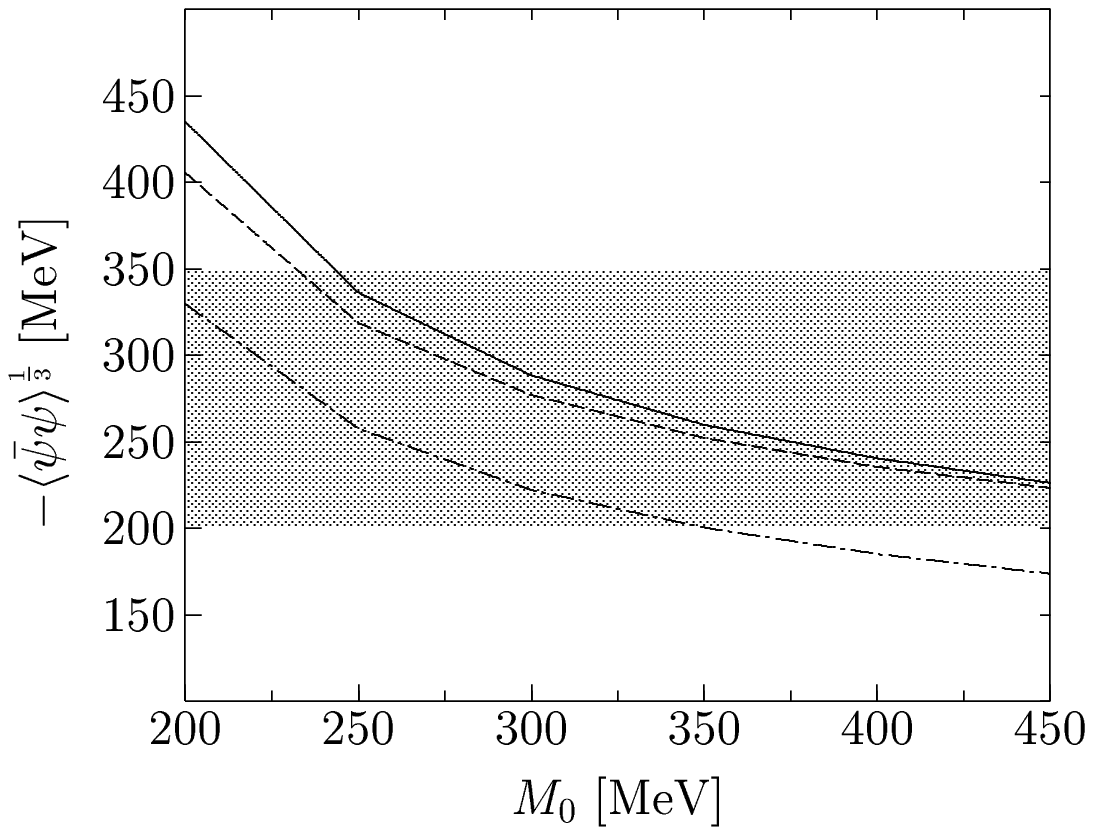}}

\vspace{0.8cm}
\noindent {\bf Fig.2}: Quark condensate as a function of $M(k)$.
The solid curve denotes the original $M(k)=M_0F^2 (k/\Lambda)$ by
Diakonov and Petrov, while the long-dashed one stands for the dipole-type
$M(k)$ and the dot-dashed one depicts the Gaussian one.  The shaded band designates
the range of the empirical values of the quark condensate.
\vspace{0.8cm}

\centerline{\includegraphics[width=15cm,height=12.0cm]{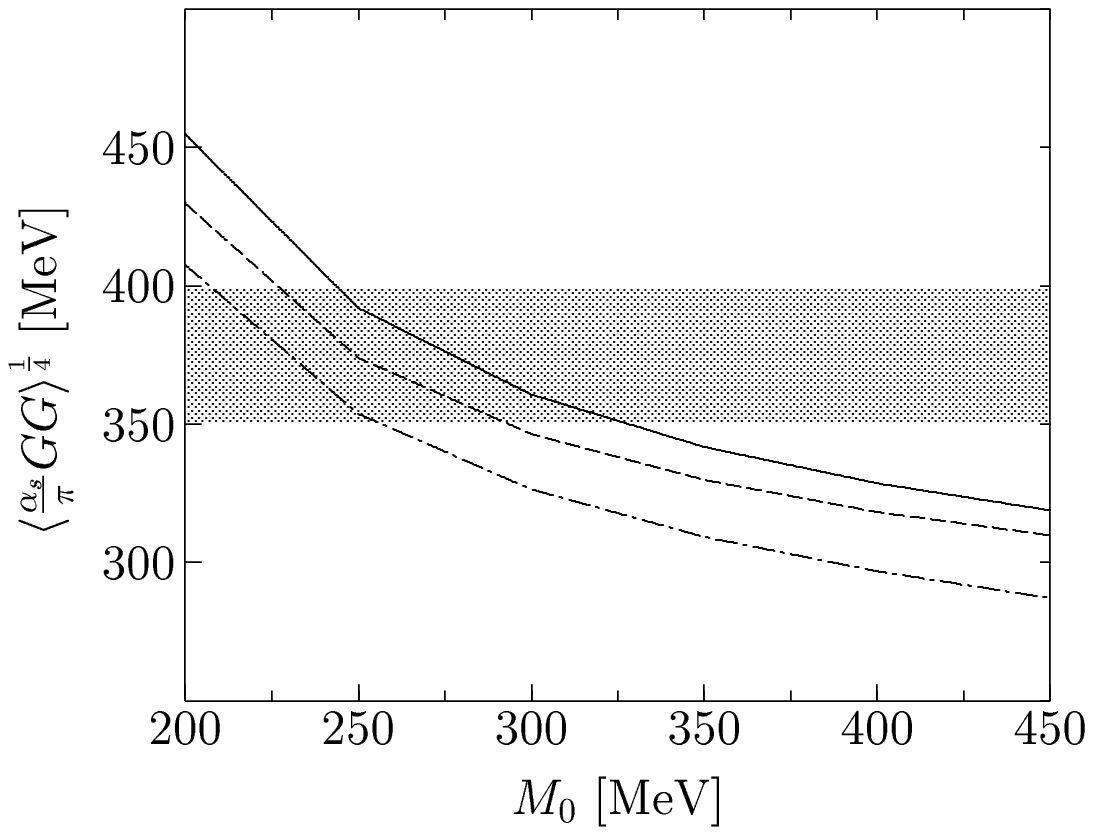}}

\vspace{0.8cm}
\noindent {\bf Fig.3}: Gluon condensate as a function of $M(k)$.
The solid curve denotes the original $M(k)=M_0F^2 (k/\Lambda)$ by
Diakonov and Petrov, while the long-dashed one stands for the dipole-type
$M(k)$ and the dot-dashed one depicts the Gaussian one.  The shaded band designates
the range of the empirical values of the gluon condensate.
\vspace{0.8cm}

\centerline{\includegraphics[width=15cm,height=12.0cm]{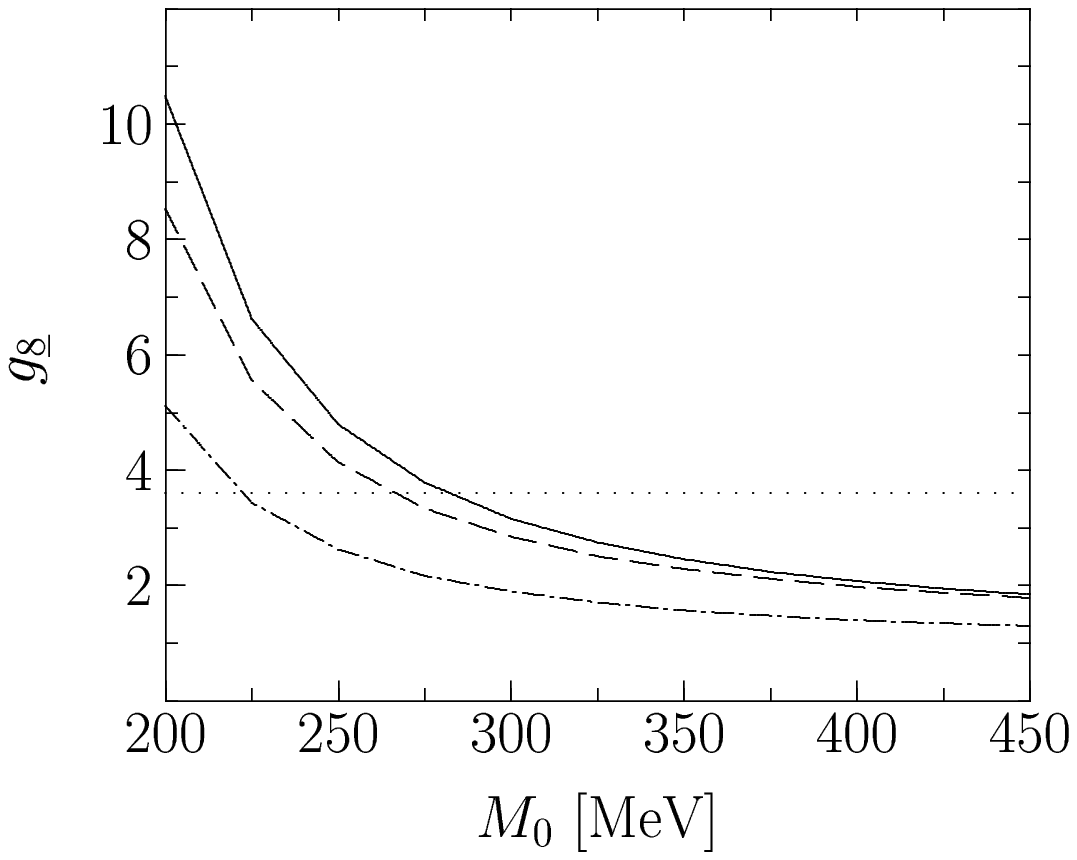}}

\vspace{0.8cm}
\noindent {\bf Fig.4}: The low energy constant $g_{\underline{8}}$ as a function of
$M(k)$.  The solid curve denotes the original $M(k)=M_0F^2 (k/\Lambda)$ by
Diakonov and Petrov, while the long-dashed one stands for the dipole-type
$M(k)$ and the dot-dashed one is the Gaussian one.  The empirical value of
$|g_{\underline{8}}|$ is approximately $3.6$ which is drawn in the dotted line.

\vspace{0.8cm}

\centerline{\includegraphics[width=15cm,height=12.0cm]{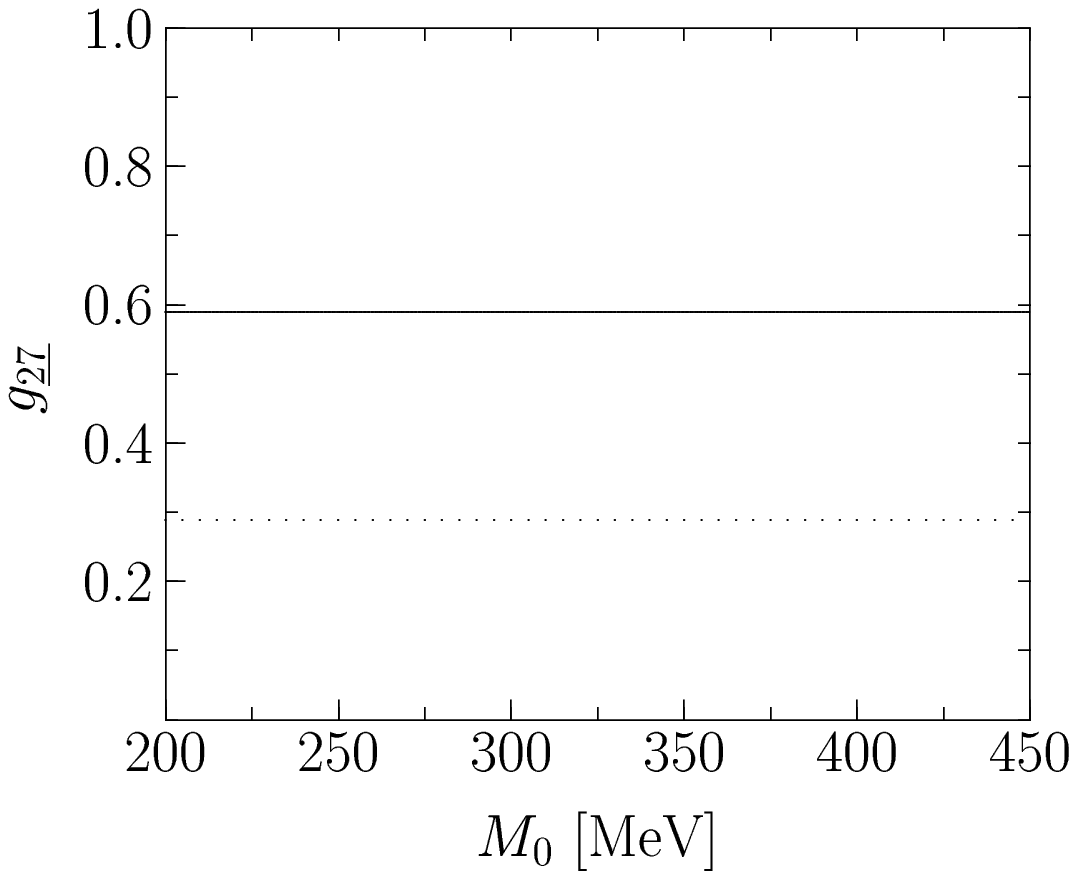}}

\vspace{0.8cm}
\noindent {\bf Fig.5}: The low energy constant $g_{\underline{27}}$ as a function of
$M(k)$.  It is independent of the $M(k)$. 

\vspace{0.8cm}

\centerline{\includegraphics[width=15cm,height=12.0cm]{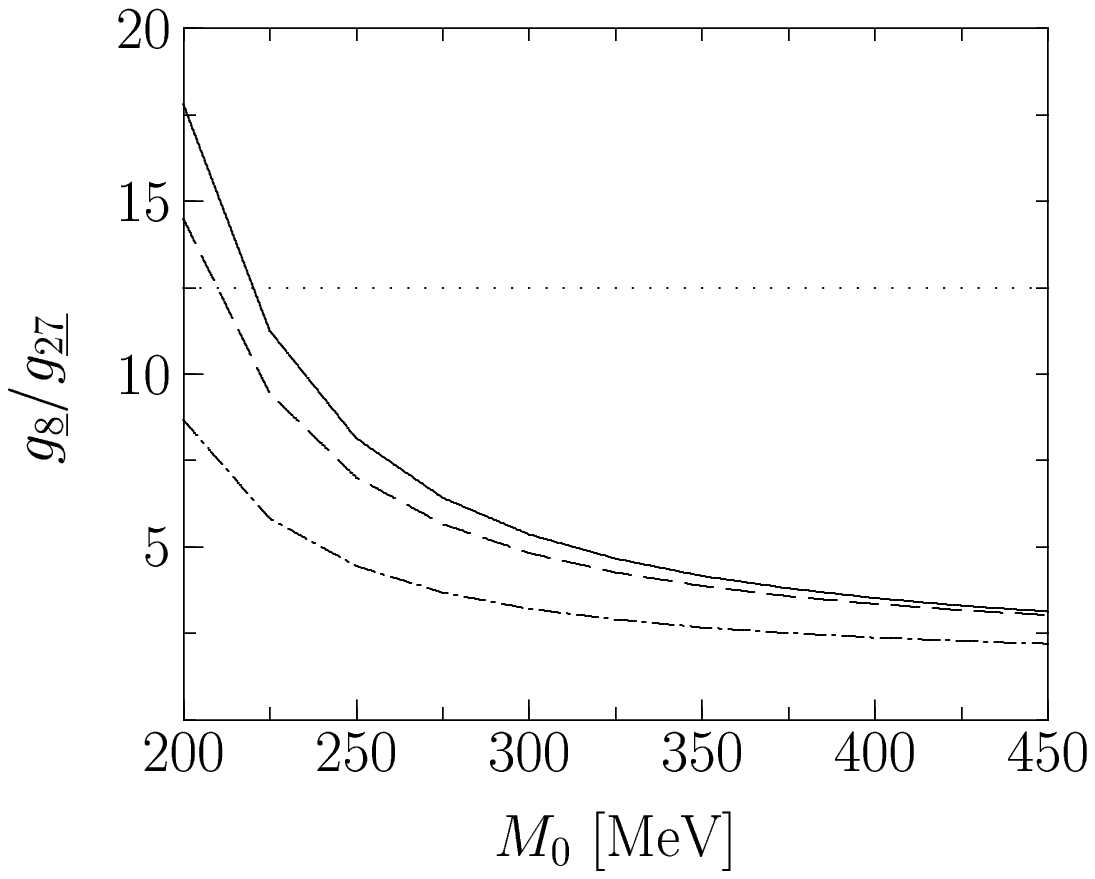}}

\vspace{0.8cm}
\noindent {\bf Fig.6}: The ratio $g_{\underline{8}}/g_{\underline{27}}$ as a function of
$M(k)$.  The solid curve denotes the original $M(k)=M_0F^2(k/\Lambda)$ by
Diakonov and Petrov, while the long-dashed one stands for the dipole-type
$M(k)$ and the dot-dashed one is the Gaussian one.  The empirical value of
$|g_{\underline{8}}/g_{\underline{27}}|$ is approximately $12.5$ which is drawn in the dotted line.

\vspace{0.8cm}

\centerline{\includegraphics[width=15cm,height=12.0cm]{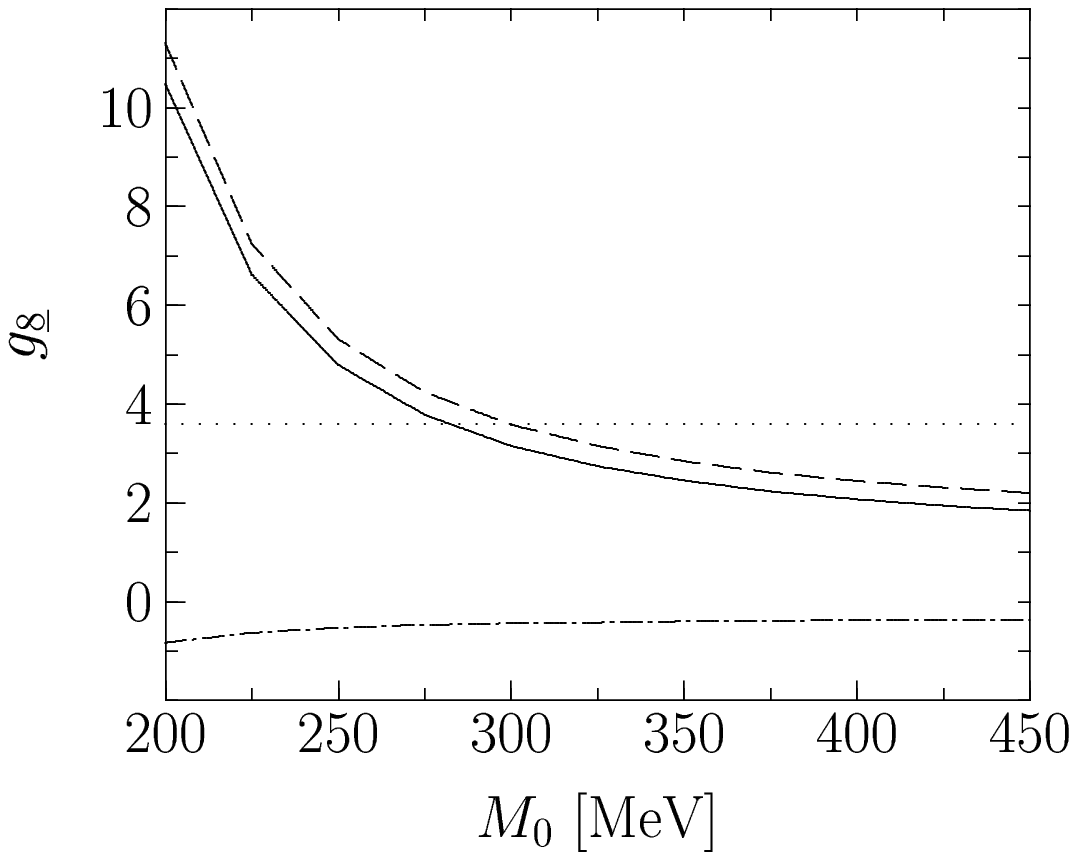}}

\vspace{0.8cm}
\noindent{\bf Fig. 7} The LO and NLO contributions to the
$g_{\underline{8}}$ in the large $N_c$ expansion.  The long-dashed
curve represents the LO contribution, the dot-dashed one the NLO
contribution, and the solid one the full result.  The empirical value of
$|g_{\underline{8}}|$ is approximately $3.6$ which is drawn in the dotted line.

\vspace{0.8cm}

\centerline{\includegraphics[width=15cm,height=12.0cm]{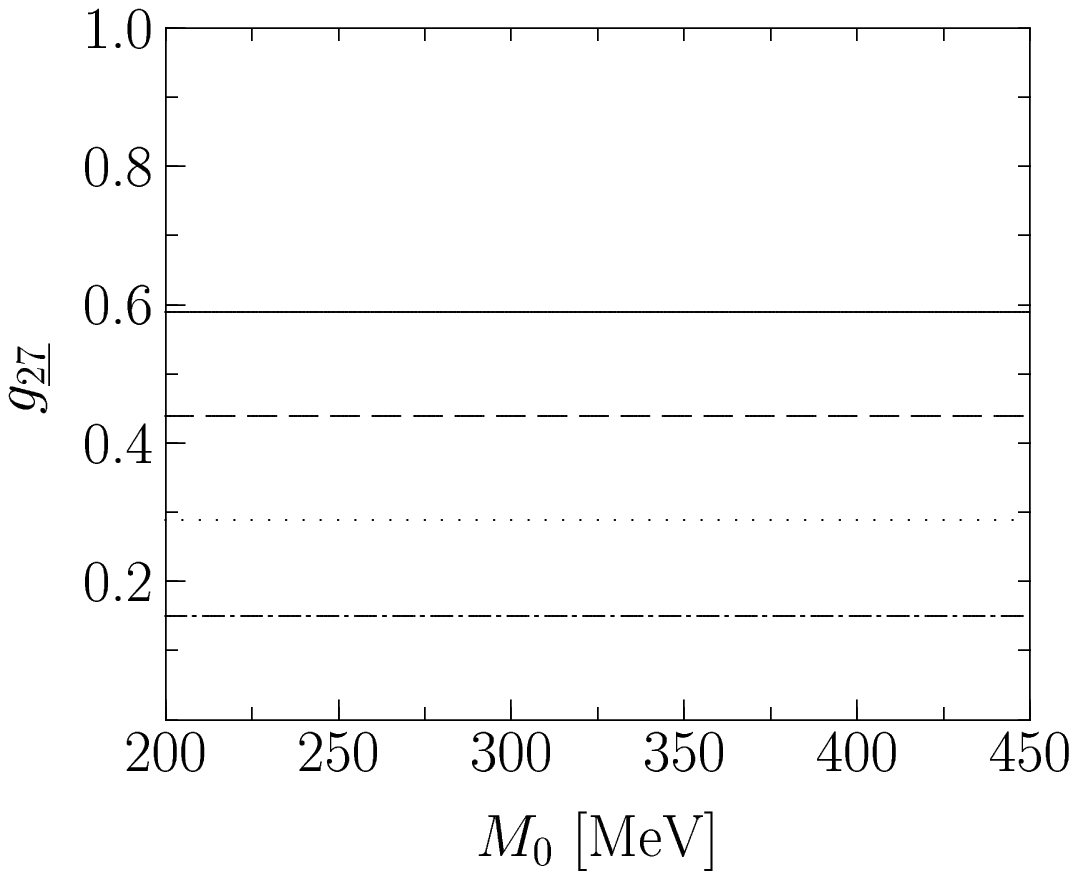}}

\vspace{0.8cm}
\noindent{\bf Fig. 8} The LO and NLO contributions to the
$g_{\underline{27}}$ in the large $N_c$ expansion.   The long-dashed
curve represents the LO contribution, the dot-dashed one the NLO
contribution, and the solid one the full result.  The empirical
value of $|g_{\underline{27}}|$ is approximately $0.29$ which is drawn in the dotted line.
\vspace{0.8cm}

\centerline{\includegraphics[width=15cm,height=12.0cm]{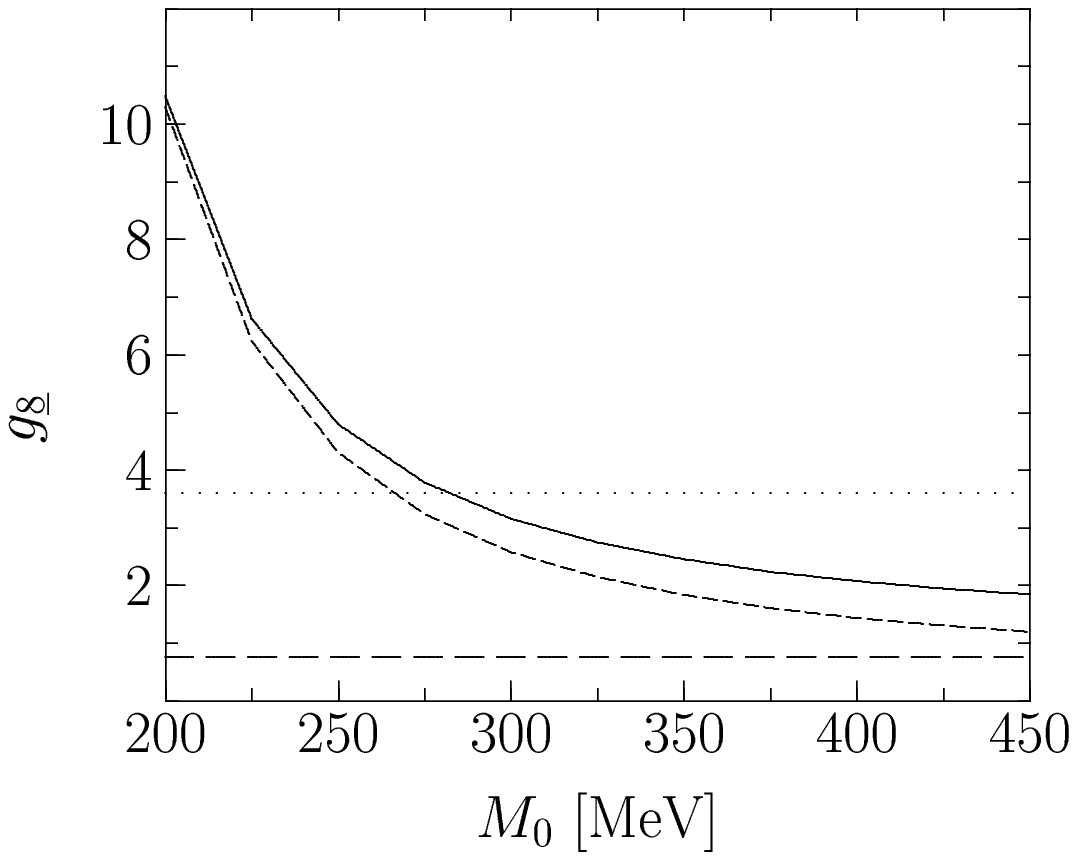}}

\vspace{0.8cm}
\noindent {\bf Fig.9}: Contributions of each quark operator $Q_i$
to the low energy constant $g_{\underline{8}}$ as a function
of $M(k)$.  The long-dashed curve denotes the contribution of the $Q_2$,
while the short-dashed one draws that of the $Q_1+Q_2$.
The dot-dashed one depicts the $g_{\underline{8}}$ with $Q_6$ added and the solid curve
represents the full result. The empirical value of $|g_{\underline{8}}|$ is approximately
$3.6$ which is drawn in the dotted line.

\vspace{0.8cm}

\end{document}